\documentclass[12pt,preprint]{aastex}

\newcommand{\q}{q_s}
\newcommand{\eff}{\epsilon}

\newcommand\lsim{\mathrel{\rlap{\lower4pt\hbox{\hskip1pt$\sim$}}
        \raise1pt\hbox{$<$}}}
\newcommand\gsim{\mathrel{\rlap{\lower4pt\hbox{\hskip1pt$\sim$}}
        \raise1pt\hbox{$>$}}}
\newcommand\propsim{\mathrel{\rlap{\lower4pt\hbox{\hskip1pt$\sim$}}
        \raise1pt\hbox{$\propto$}}}

\newcommand{\yr}{\,\mathrm{yr}}

\begin{document}

\title{The Population of Viscosity-- and Gravitational Wave--Driven Supermassive Black Hole Binaries Among Luminous AGN}

\author{Zolt\'an Haiman,\altaffilmark{1} Bence Kocsis,\altaffilmark{2,3,4} and Kristen Menou,\altaffilmark{1}}

\affil{$^1$Department of Astronomy, Columbia University, 550 W120th St., New York, NY 10027}

\affil{$^2$Harvard-Smithsonian Center for Astrophysics, 60 Garden St., Cambridge, MA 02138}

\affil{$^3$Institute for Advanced Study, Einstein Dr., Princeton, NJ 08540}

\affil{$^4$Institute of Physics, E\"otv\"os University, P\'azm\'any P. s. 1/A, 1117 Budapest, Hungary}

\begin{abstract}
  Supermassive black hole binaries (SMBHBs) in galactic nuclei are
  thought to be a common by--product of major galaxy mergers. We use
  simple disk models for the circumbinary gas and for the binary-disk
  interaction to follow the orbital decay of SMBHBs with a range of
  total masses ($M$) and mass ratios ($q$), through physically
  distinct regions of the disk, until gravitational waves (GWs) take
  over their evolution.  Prior to the GW--driven phase, the viscous
  decay is generically in the stalled ``secondary--dominated'' regime.
  SMBHBs spend a non--negligible fraction of a fiducial time of $10^7$
  years at orbital periods between days $\lsim t_{\rm orb}\lsim$ year,
  and we argue that they may be sufficiently common to be detectable,
  provided they are luminous during these stages. A dedicated optical
  or X--ray survey could identify coalescing SMBHBs statistically, as
  a population of periodically variable quasars, whose abundance obeys
  the scaling $N_{\rm var}\propto t_{\rm var}^{\alpha}$ within a range
  of periods around $t_{\rm var}\sim$ tens of weeks.  SMBHBs with $M
  \lsim 10^7 \,{\rm M_\odot}$, with $0.5\lsim \alpha\lsim 1.5$, would
  probe the physics of viscous orbital decay, whereas the detection of
  a population of higher--mass binaries, with $\alpha=8/3$, would
  confirm that their decay is driven by GWs.  The lowest mass SMBHBs
  ($M \lsim 10^{5-6} \,{\rm M_\odot}$) enter the GW-driven regime at
  short orbital periods, when they are already in the frequency band
  of the {\it Laser Interferometric Space Antenna} ({\it LISA}). While
  viscous processes are negligible in the last few years of
  coalescence, they could reduce the amplitude of any unresolved
  background due to near--stationary {\it LISA} sources.  We discuss
  modest constraints on the SMBHB population already available from
  existing data, and the sensitivity and sky coverage requirements for
  a detection in future surveys.  SMBHBs may also be identified from
  velocity shifts in their spectra; we discuss the expected abundance
  of SMBHBs as a function of their orbital velocity.
\end{abstract}

\vspace{\baselineskip}


\vspace{\baselineskip}

\section{Introduction}

Supermassive black holes (SMBHs) appear to be present in the nucleus
of most, and perhaps all, nearby galaxies (see, e.g., reviews by
\citealt{kr95} and \citealt{ff05}).  The correlations between the
masses of the SMBHs and various global properties of the host galaxies
suggest that evolution of SMBHs is closely related to the evolution of
galaxies.  In particular, in hierarchical structure formation models,
galaxies are built up by mergers between lower--mass progenitors.
Each merger event is expected to deliver the nuclear SMBHs
(e.g. \citealt{springel05,robertson06}), along with a significant
amount of gas \citep{bh92}, to the central regions of the new
post--merger galaxy.

There is some evidence for nuclear supermassive black hole binaries
(SMBHBs), which would be expected to be produced in galaxy mergers.
Direct X--ray imaging of an active nucleus \citep{komossa03} has
revealed a SMBH binary at a separation of $\sim1$kpc, and
\citet{boroson09} recently identified a candidate SMBHB, at $\sim
10^4$ times smaller separation, from its optical spectrum.  A radio
galaxy is also known to have a double core with a projected separation
of $\sim 10$ pc \citep{rodriguez06}, and several other observations of
radio galaxies, such as the wiggled shape of jets indicating
precession \citep[e.g.][]{roos}, the X--shaped morphologies of radio
lobes \citep[e.g.][]{merritt,liu04}, the interruption and recurrence
of activity in double--double radio galaxies
\citep[e.g.][]{schoenmakers,liu03}, and the elliptical motion of the
unresolved core of 3C66B \citep{sudou}\footnote{The lack of any
modulation in arrival times for radio pulsars suggests that the
elliptical motion of the last source has a different origin; see
\S~\ref{sec:existing} below.}  have all been interpreted as indirect
evidence for SMBH binaries down to sub--pc scales.

Two interesting conclusions may be inferred from the above
observations.  First, while there is evidence for a handful of nuclear
SMBHBs, these objects appear to be rare. This suggests that if
binaries do form frequently, then they coalesce (or at least their
orbital separation decays to undetectably small values) in a small
fraction of the Hubble time.  Second, SMBHBs can apparently produce
bright emission, with a luminosity comparable to active galactic
nuclei (AGN), before they coalesce.  In general, the circumbinary gas,
delivered to the nucleus in galactic mergers, can both play a catalyst
role in driving rapid SMBHB coalescence \citep{bbr80,gr00,elcm04}, and
could also accrete onto one or both SMBHs, accounting for bright
emission during the orbital decay.

The dense nuclear gas around the BH binary is expected to cool
rapidly, and settle into a rotationally supported, circumbinary disk
\citep[e.g.][]{barnes,elcm05}.  The dynamical evolution of a SMBHB
embedded in such a thin disk has been studied in various idealized
configurations
\citep[e.g.][]{an02,liu03,mp05,dot06,mm08,hayasaki09,cuadra09}.  The
generic conclusion of these studies is that initially, the orbital
decay is relatively slow, and is dominated by viscous angular momentum
exchange with the gas disk, whereas at small separations, the decay is
much more rapid, and is eventually dominated by gravitational wave
(GW) emission.

Whether the decaying SMBHBs produce bright electromagnetic (EM)
emission is comparatively much less well understood.  If the disk is
thin, the torques from the binary create a central cavity, nearly
devoid of gas, within a region about twice the orbital separation
\citep[for a nearly equal--mass binary, e.g.][]{al94}, or a narrower
gap around the orbit of the lower--mass BH in the case of unequal
masses $q \equiv M_2/M_1 \ll 1$ \citep[e.g.][]{an02}.  In the latter
case, the lower--mass hole ``ushers'' the gas inward as its orbit
decays, producing a prompt and luminous signal during coalescence.  In
the former case, if the central cavity were indeed truly empty, no gas
would reach the SMBHBs, and bright emission could not be produced.
However, numerical simulations suggest residual gas inflow into the
cavity \citep{al96,mm08,hayasaki07,hayasaki08,cuadra09}, which may
plausibly accrete onto the BHs, producing non--negligible EM
emission.\footnote{This would be followed by an X--ray ``afterglow''
$\sim 7(1+z)(M/10^6{\rm M_\odot})^{1.32}$ yr after the coalescence,
caused by the gas outside the cavity falling in, after a delay set by
the disk viscous time \citep{liu03,mp05}.  Such an afterglow is
interesting, for example for a follow--up to SMBH merger events
detected by the {\it Laser Interferometric Space Antenna (LISA)}, but
not relevant to the idea proposed in the present {paper}.}  Finally,
SMBHBs recoil at the time of their coalescence due to the emission of
gravitational waves \citep{blanchet05}.  The gas disk will respond
promptly (on the local orbital timescale) to such a kick, which may
produce shocks, and transient EM emission, after
coalescence~\citep{lippaidisk,schnittmankick,shieldskick}.  The kick,
however, can begin building up during the late inspiral phase
\citep{anatomy08}, possibly resulting in some emission even before the
final coalescence. During the late stages of coalescence, emission may
also be produced by viscous heating of the disk by the GWs themselves
\citep{kocsisloeb08}.

The luminosity, spectrum, and time--evolution of any EM emission
produced by coalescing SMBHBs, especially during the last, GW--driven
stages, remains uncertain.  {\it However, any emission produced during
  the inspiral stage is likely to be variable.}  For example, recent
numerical simulations of an equal--mass binary on parsec scales
\citep{mm08}, and of both equal and unequal--mass binaries on
sub--parsec scales \citep{hayasaki07,hayasaki08,cuadra09} find that
the circumbinary gas disk is perturbed into eccentric orbits by the
rotating quadrupole potential of the binary, and that the rate of
residual accretion across the edge of the cavity is modulated,
tracking the orbital period.  The luminosity is likely to be directly
tied to the mass accretion rate, and therefore may vary periodically.
However, even if the gas accretion rate were steady, one would expect
periodic flux variations, due to the orbital motion of the binary
(Kocsis \& Loeb, in preparation).

In this paper, we address the question: {\it Given their expected rate
  of orbital decay, could the population of coalescing SMBHBs be
  identified statistically in an observational survey for periodically
  variable sources?}  Given that the interpretation of individual
SMBHB candidates have so far remained ambiguous, with alternative
explanations possible for each source, the potential for such a
statistical identification should be explored.

To answer this question, we first utilize steady--state thin disk
models to study the orbital decay of a SMBHB, embedded in a
circumbinary disk.  The decay is described by the residence time
$t_{\rm res}\equiv -R(dR/dt)^{-1}$ the sources spend at each orbital
radius $R$, or at the corresponding orbital timescale $t_{\rm orb}$.
In the limiting case of a purely GW--driven evolution, which becomes
valid at small orbital separations (typically at $\lsim$
several$\times 10^2$ Schwarzschild radii, but with large variations;
see below) and remains valid until the final minutes of the merger
(the so--called ``plunge'' stage), the residence time is given by
$t_{\rm res}=t_{\rm GW}\propto t_{\rm orb}^{8/3}$.  At larger
separations, the viscous interaction between the binary and the disk
drives the binary evolution. The residence time in this regime becomes
dependent on assumptions about the properties of the disk and the
nature of the binary--disk interaction, which we will explore in this
paper. In general, $t_{\rm res}\propto t_{\rm var}^\alpha$, with the
generic value of $\alpha$ well below $8/3$ -- significantly flatter
than the $t_{\rm res}\, vs.\, t_{\rm orb}$ relation in the GW--driven
stage.

We then hypothesize that (i) non--negligible emission (at a fair
fraction of the Eddington luminosity) is maintained throughout the
orbital decay, and (ii) the luminosity varies periodically on the
orbital time--scale.  The first assumption allows us to identify
coalescing SMBHBs with luminous quasars.  The second assumption
implies that as the orbit of a binary decays, its variability
timescale decreases.  Among sources at redshift $z$ with similar
inferred BH masses, the observed incidence rate $f_{\rm var}$ of
periodic variability on the time--scale $t_{\rm var}\sim (1+z)t_{\rm
  orb}$, is then proportional to the residence time $t_{\rm
  res}=t_{\rm res}(t_{\rm var})$.  At short periods, the $f_{\rm var}$
could therefore show a characteristic power--law dependence on $t_{\rm
  var}$ indicative of a GW--driven evolution, whereas at longer
periods (and, as we will discuss, for lower BH masses) the dependence
will be flatter, due to viscosity--driven
evolution.\footnote{Throughout this manuscript, we will use the term
  ``viscosity--driven evolution'' to refer to the exchange of angular
  momentum and energy in the binary--disk system that arises from the
  combination of gas viscosity and the tidal torques from the binary.}

We quantify the requirements that such periodically variable sources
be identifiable, based on their incidence rate, in an optical
or X--ray survey.  Luminosity variations at a fraction $f_{\rm
Edd}\lsim 0.01$ of the Eddington luminosity would correspond to a
periodically varying flux component with amplitude $(f_{\rm Edd}/0.01)
(M_{\rm bh}/3\times 10^7{\rm M_\odot}) 10^{-15}~{\rm
erg~s^{-1}~cm^{-2}}$ for BHBs at $z=2$, or to $i\approx
26+2.5\log[(f_{\rm Edd}/0.01) (M_{\rm BH}/3\times 10^7{\rm
M_\odot})^{-1}]$ magnitudes in the optical.  We find that these
periodic sources are either too faint or too rare to have been found
in existing variability surveys.  However, if the overall luminosity
is indeed a non--negligible fraction of the binary's Eddington
luminosity, then a long--duration future survey, sensitive to periods
of weeks to tens of weeks, could look for periodically variable
sources, and identify a population of sources obeying well--defined
scaling laws.

The discovery of a population of such periodically variable sources
could have several implications.  At long periods and low BH masses,
the scaling index $\alpha$ between the residence time and the period
$t_{\rm res} \propto t_{\rm orb}^\alpha$ will probe the physics of the
circumbinary accretion disk and viscous orbital decay.  At shorter
periods and higher masses (roughly, at $t_{\rm orb}<$ few weeks for
$M>10^7~{\rm M_\odot}$), the identification of a $f_{\rm var}\propto
t_{\rm orb}^{8/3}$ power--law would confirm that the orbital decay is
driven by GWs.  This would amount to an indirect, statistical
detection of GW--driven SMBHBs, independent of any direct detection of
GWs by {\it LISA}.  This would also confirm that circumbinary gas is
present at small orbital radii and is being perturbed by the BHs --
and would thus serve as a proof of concept for finding {\it LISA}
electromagnetic counterparts.

The rest of this paper is organized as follows.
In \S~\ref{sec:diskmodels}, we discuss the evolution of binaries with
different masses and mass-ratios, embedded in a circumbinary gas disk.
We describe simplified models for the disk and for the binary--disk
interaction, and emphasize that the binaries probe the distinct
physical regimes in the disk, before GWs take over their evolution.
In \S~\ref{sec:observations}, we discuss the possibility of searching
for a population of coalescing SMBHBs among a catalog of luminous
quasars, either based on their variability, or on shifts of their
spectral lines.  We discuss modest constraints available from existing
surveys, and comment on specific recently detected individual SMBHB
candidates.  We then quantify the requirements for a detection in a
future survey.
In \S~\ref{sec:summary}, we briefly summarize our results and offer our
conclusions.
When necessary in this paper, we adopt the background cosmological
parameters $\Omega_m=0.3$, $\Omega_{\Lambda}=0.7$, and $H_0=70~{\rm
km~s^{-1}~Mpc^{-1}}$ \citep{wmap}.

\section{Binary Evolution}
\label{sec:diskmodels}

In this section, we describe the evolution of the orbital separation
of a SMBH binary. The basic picture we adopt is that the binary is
embedded in a thin circumbinary disk, with the plane of the disk
aligned with the binary's orbit \citep{bp75,ivanov}.  Initially, the
orbital decay is dominated by viscous angular momentum exchange with
the gas disk.  However, the time--scale for viscous decay decreases
relatively slowly as the orbital separation $R$ decreases ($t_{\rm
res}\propto R^{1/2}-R^{11/4}$; see below) whereas the time--scale to
decay due to gravitational radiation decreases steeply ($t_{\rm
GW}\propto R^4$).  Therefore, generically, there exists a critical
orbital radius $R_{\rm crit}$, below which the decay is dominated by
gravitational radiation.

To describe the evolution quantitatively, we make several simplifying
assumptions. The circumbinary gas is assumed to form a standard
geometrically thin, optically thick, radiatively efficient,
steady--state accretion disk \citep{ss73}.  We assume zero
eccentricity for both the binary and for the disk (justified by
\citealt{dot06}, however see
\citealt{an05,mm08,dot08,hayasaki07,cuadra09}), and we assume
co--planarity between the disk and the binary \citep{bp75,ivanov}.
All of these assumptions may fail in the late stages of the merger
(even before GW--driven decay begins).  However, under these
assumptions, the disk structure and the orbital decay have simple
limiting power--law solutions, with the power--law indices depending
on the choice for the underlying physics.  These solutions are useful
to describe the possible evolution of the binary, and to illustrate
the point that the decay rate is generically a different -- much
flatter -- function of $t_{\rm orb}$ than the $t_{\rm GW}\propto
t_{\rm orb}^{8/3}$ behavior in the GW--driven case.

We emphasize that our aim here is not to provide accurate,
self--consistent solutions for the co--evolution of the SMBH binary
and circumbinary disk.  Rather, we derive only gross scaling laws in
various regimes -- our main point is that these regimes and associated
uncertainties, which are large, can in principle be probed
observationally.

\subsection{Notation}

We adopt the following notation throughout this paper.  We refer the
reader to \citet{shapiroteukolsky} and \citet{accretionpower} for
general introductions to accretion disks.

\begin{itemize}

\item{\em Physical constants:}
$\rm G$ is the gravitational constant;
$\rm c$ is the speed of light;
$\rm{k_B}$ is the Boltzmann constant;
$\sigma_{\rm SB}$ is the Stefan--Boltzmann constant;
$\sigma_{\rm T}$ is the Thompson cross section;
$\mu_e=n_e m_{\rm H}/\rho$ is the mean mass per electron in units of hydrogen atom mass, $m_{\rm H}$, which satisfies $\mu_e=(1+X_{\rm H})/2$ for a fully ionized gas of both hydrogen and helium;
$X_{\rm H}$ is the mass fraction of hydrogen;
$\mu_0=2/(3X_{\rm H}+1)$ is the mean molecular weight;
$\kappa_{\rm es}=\mu_e\sigma_{\rm T}/m_{\rm H}$ is the electron scattering opacity;
and $\kappa_{\rm ff}=(8\times 10^{22} {\rm cm}^2\,{\rm g}^{-1}) \mu_e[\rho/({\rm g}\,{\rm cm}^{-3})] (T/{\rm K})^{-7/2}$
is the Rosseland mean absorption opacity in the free--free regime
\citep{padmanabhan,rybicki,shapiroteukolsky}.

\item{\em BH parameters:}
$M_1$ and $M_2$ are the individual BH masses;
$M=M_1+M_2$ is the total BH mass;
$q=M_2/M_1\leq 1$ is the mass ratio;
$\q = 4q/(1+q)^2$ is the normalized symmetric mass ratio;
$\mu=\q M/4$ is the reduced mass;
$R$ is the binary separation;
$R_0=R/(1+q)$ is the location of the lower--mass secondary, measured from the center of mass of the binary;
$R_S=2{\rm G}M/{\rm c}^2$ is the Schwarzschild radius corresponding to the total mass;
$L_{\rm Edd}=4\pi {\rm G}{\rm c}\kappa_{\rm es}^{-1} M$ is the Eddington luminosity for a BH of mass $M$;
$\dot M_{\rm Edd}\equiv L_{\rm Edd}/(\eff {\rm c}^2)$ is the Eddington accretion rate with a radiative efficiency $\eff$; and $t_{\rm Edd}\equiv M/\dot M_{\rm Edd} = \kappa_{\rm es}\eff c/(4\pi {\rm G}) = (3.94\times 10^7\yr)\times\tilde\mu_e \eff_{0.1} $ is the characteristic time--scale associated with Eddington accretion.

\item{\em Disk parameters:}
$H$ is the vertical scale height (the effective geometrical semi--thickness of the disk);
$\rho$ is the volumic gas density;
$\Sigma=\rho/(2 H)$ is the surface density;
$P_{\rm gas}$ is the gas pressure;
$P_{\rm rad}$ is the radiation pressure;
$P=P_{\rm rad}+P_{\rm gas}$ is the total pressure;
$\beta \equiv P_{\rm gas}/(P_{\rm rad}+P_{\rm gas})$;
$T$ is the (midplane) gas temperature;
$T_{\rm eff}$ is the effective temperature defined such that the locally emitted flux through an infinitesimal disk surface element is $\sigma_{\rm SB} T_{\rm eff}^{4}$;
$\Omega(r)$ is the Keplerian orbital angular velocity;
$R_{\lambda}$ is the outer radius of the gap in the punctured circumbinary disk, measured from the center of mass of the binary;
$\eta$ is the anomalous dynamical viscosity;
$\nu=\eta/\rho$ is the anomalous kinematic viscosity;
$\alpha$ is the standard viscosity parameter of thin accretion disks;
$b$ is a constant, either 0 or 1, determining whether viscosity scales with the total or just the gas pressure, so that $\eta \equiv \alpha P \beta^b\Omega^{-1}$;
$\kappa$ is the opacity of the disk material;
$\tau= (1/2)\kappa\Sigma$ is the vertical optical depth;
and $f_{T}$ is a constant defined such that $f_{T} =\tau^{-1} T^4/T_{\rm eff}^4$.
Quantities with a $\lambda$ subscript (e.g. $\Omega_\lambda,
\Sigma_\lambda, H_\lambda$) denote parameters in a steady--state disk
around a single unperturbed accreting BH, computed at the radius
$R_{\lambda}$.  Similarly, quantities with a 0 subscript are evaluated
at the position of the secondary $R_0$.

\end{itemize}

With the above definitions, we proceed to define the dimensionless
quantities $r=R/R_S$, $r_3=r/10^3$, $M_7=M/(10^7 M_{\odot})$, $\dot m
= \dot M/\dot M_{\rm Edd}$, $\dot m_{0.1}=\dot m/0.1$,
$\alpha_{0.3}=\alpha/0.3$, $\eff_{0.1}=\eff/0.1$, $\lambda=
R_{\lambda}/R$, $\tilde\kappa_{\rm es}=\kappa/\kappa_{{\rm es}}$, and
$\tilde\kappa_{\rm ff}=\kappa/\kappa_{\rm ff}$.  Note that radii are
measured from the center of mass of the binary.  We adopt the set of
fiducial values $X_{\rm H}=0.75$, $\mu_e=0.875$, $\mu_0=0.615$,
$f_T=3/4$,
\footnote{The choice $f_{T}=3/4$ is appropriate for a one--zone model
where all the energy is dissipated near the midplane and the opacity
is constant vertically
\citep{shapiroteukolsky,accretionpower,arm07}. For reference, we note
that \citet{goodman03} and \citet{sirkogoodman03} adopt different
values of $f_{T}=1$ and $f_{T}=3/8$, respectively.}
and denote values relative to these fiducial choices with a tilde,
e.g. $\tilde\mu_e=\mu_e/0.875$. Our fiducial binary+disk model is
therefore chosen to be $\tilde\mu_e=\tilde\mu_0=\tilde
f_T=q=\q=M_7=\dot m_{0.1}=\eff_{0.1}=\alpha_{0.3}=\lambda=1$.  Since
all of our expressions can be written as products of power--laws in
the physical parameters, the resulting expressions become tractable in
these units.

\subsection{Thin Disk Models}
\label{subsec:thindisks}

We next collect the basic expressions from the literature for
accretion disk models under different physical conditions.  We quote
the equations for a range of different steady thin disks, valid for a
single accreting BH \citep{ss73}. We distinguish several cases: {\it
  (i)} whether the radiation or gas pressure provides the dominant
vertical support, {\it (ii)} whether the opacity is dominated by
electron scattering, $\kappa_{\rm es}$, or free--free absorption,
$\kappa_{\rm ff}$, and {\it (iii)} whether the viscosity $\eta$ is
proportional to the total pressure or the gas pressure (also known as
$\alpha$ and $\beta$ disk models, respectively).  Based on these
choices, the accretion disk can be divided radially into three
distinct regions \citep{shapiroteukolsky}:

\begin{enumerate}

\item {\it Inner region:} Radiation pressure and electron-scattering
  opacity dominate, $P\approx P_{\rm rad}$, $\tilde\kappa_{\rm
    es}\approx 1$, valid inside $r_3\ll r_3^{{\rm gas}/{\rm rad}}$
  where $r_3^{{\rm gas}/{\rm rad}}$ is defined in
  equations~(\ref{e:gas/radb1}) and~(\ref{e:gas/radb0}) below.

\item {\it Middle region:} Gas pressure and electron-scattering
  opacity dominate, $P\approx P_{\rm gas}$, $\tilde\kappa_{\rm
  es}\approx 1$, valid between $ r_3^{{\rm gas}/{\rm rad}}\ll r_3
  \lsim r_3^{{\rm es}/{\rm ff}}$, where $r_3^{{\rm es}/{\rm ff}}$ is
  defined in equation~(\ref{e:es/ff}) below.

\item {\it Outer region:} Gas pressure and free-free opacity dominate,
  $P\approx P_{\rm gas}$, $\tilde\kappa_{\rm ff}\approx 1$, valid
  outside of $r_3 \gsim r_3^{{\rm es}/{\rm ff}}$.

\end{enumerate}

In region (1), it makes a difference whether the viscosity is
proportional to the total pressure or just the gas pressure, labeled
below by $b=0$ or 1, (i.e. $\alpha$ or $\beta$ disk) respectively. In
all cases, we assume that the disk is optically thick, i.e. $\tau \gg
1$. We obtain $\Sigma(r)$ and $H(r)$ following \citet{goodman03} or
\citet{goodmantan04},
\begin{eqnarray}\label{e:Sigma_definition}
 \Sigma(r) &=& \frac{2^{4/5}}{3\pi^{3/5}} \sigma_{\rm SB}^{1/5} \left(\frac{\mu_0 m_{\rm H}}{{\rm k_B}}\right)^{4/5} f_{T}^{-2}\alpha^{-4/5} \kappa^{-1/5} \dot M^{3/5} \Omega^{2/5} \beta^{-(4/5)(b-1)},\\\label{e:H_definition}
 H(r) &=& \frac{f_{T} \kappa \dot M}{2\pi {\rm c} (1-\beta)}.
\end{eqnarray}
where $b=0$ or 1, and the radial dependence is implicit in $\Omega$ and $\beta$.
Here, $\beta(r) \equiv P_{\rm gas}/(P_{\rm rad}+P_{\rm gas})$ which satisfies
\begin{eqnarray}
  \frac{\beta^{(1/2) + (1/10)(b-1)}}{1-\beta} &=& 2^{3/5}\pi^{4/5} {\rm c} \sigma_{\rm SB}^{-1/10} \left(\frac{{\rm k_B}}{\mu_0 m_{\rm H}}\right)^{2/5}\alpha^{-1/10} \kappa^{-9/10}\dot M^{-4/5} \Omega^{-7/10} \label{e:beta_definition}.
\end{eqnarray}

The asymptotic limits of equations~(\ref{e:Sigma_definition}) and
(\ref{e:H_definition}) can be obtained in regions (1--3), using
equation~(\ref{e:beta_definition}). The results are

\noindent{\em Inner region:}
\begin{eqnarray}\label{e:Sigma_inb1}
 \Sigma(r) &=&  (1.63 \times 10^5 {\,\rm g}{\,\rm cm}^{-2})  \mu_0^{4/5}\mu_e^{-4/5}\tilde\kappa_{\rm es}^{-1/5}  f_{T}^{-2} \alpha_{0.3}^{-4/5}  \left(\frac{\dot m}{\eff_{0.1}}\right)^{3/5} M_7^{1/5} r_3^{-3/5} \rm{~~if~} b=1,\\\label{e:Sigma_inb0}
           &=& (2.50 \times 10^4 {\,\rm g}{\,\rm cm}^{-2}) \mu_e^{-1} \tilde\kappa_{\rm es}^{-2}  f_{T}^{-2} \alpha_{0.3}^{-1} \left(\frac{\dot m}{\eff_{0.1}}\right)^{-1}  r_3^{3/2} \rm{~~~~if~} b=0, \\\label{e:H_in}
 H(r) &=&  (10.0  R_S)  f_{T} \frac{\dot m}{\epsilon_{0.1}}~~~~~{\rm for}~{\rm arbitrary}~b.
\end{eqnarray}

\noindent{\em Middle region:}
\begin{eqnarray}\label{e:Sigma_middle}
 \Sigma(r) &=& (1.63 \times 10^5 {\,\rm g}{\,\rm cm}^{-2})  \mu_0^{4/5}\mu_e^{-4/5} \tilde\kappa_{\rm es}^{-1/5} f_{T}^{-2} \alpha_{0.3}^{-4/5}  \left(\frac{\dot m}{\eff_{0.1}}\right)^{3/5} M_7^{1/5} r_3^{-3/5}, \\\label{e:H_middle}
 H(r) &=& (3.11  R_S)  \mu_e^{-1/10} \mu_0^{-2/5}   \tilde\kappa_{\rm es}^{1/10}f_{T} \alpha_{0.3}^{-1/10} \left(\frac{\dot m}{\epsilon_{0.1}}\right)^{1/5} M_7^{-1/10} r_3^{21/20}.
\end{eqnarray}

\noindent{\em Outer region:}
\begin{eqnarray}\label{e:Sigma_out}
 \Sigma(r) &=& (2.61 \times 10^5 {\,\rm g}{\,\rm cm}^{-2}) \mu_e^{-4/5} \mu_0^{3/4} \tilde\kappa_{\rm ff}^{-1/10}  f_{T}^{-143/80} \alpha_{0.3}^{-4/5} \left(\frac{\dot m}{\eff_{0.1}}\right)^{7/10} M_7^{1/5} r_3^{-3/4}, \\\label{e:H_out}
 H(r) &=& (3.08  R_S) \mu_e^{-1/10}\mu_0^{-3/8}\tilde\kappa_{\rm ff}^{1/20}  f_{T}^{143/160} \alpha_{0.3}^{-1/10}\left(\frac{\dot m}{\epsilon_{0.1}}\right)^{3/20} M_7^{-1/10} r_3^{9/8}.
\end{eqnarray}

The boundaries between the inner/middle and middle/outer regions can
be found from equations
(\ref{e:Sigma_definition})-(\ref{e:beta_definition}), by requiring
$P_{\rm gas}=P_{\rm rad}$ and $\kappa_{\rm ff}=\kappa_{\rm es}$,
respectively. Note that $\kappa_{\rm ff}(r)\propto \rho T^{7/2}$
depends on radius implicitly through the density and the
temperature. Using the (mid-plane) temperature given by
\citet{goodmantan04},
\begin{eqnarray}\label{e:Temperature_definition}
 T(r) &=& \left(16\pi^2\right)^{-1/5}
\left(\frac{\mu_0 m_{\rm H}}{{\rm k_B}\sigma_T}\right)^{1/5}
\alpha^{-1/5} \kappa^{1/5} \dot M^{2/5} \Omega^{3/5} \beta^{-(1/5)(b-1)},
\end{eqnarray}
we find that the transitions are located at the radii
\begin{eqnarray}
r_3^{{\rm gas}/{\rm rad}} &=& 0.482 \,\tilde\mu_0^{8/21}\tilde\mu_e^{2/21} \tilde\kappa_{\rm es}^{6/7} \alpha_{0.3}^{2/21} (\dot m_{0.1}/\epsilon_{0.1})^{16/21} M_7^{2/21}~~\rm{~~if~} b=1,\label{e:gas/radb1}\\
                          &=& 0.515 \,\tilde\mu_0^{8/21}\tilde\mu_e^{2/21} \tilde\kappa_{\rm es}^{6/7} \alpha_{0.3}^{2/21} (\dot m_{0.1}/\epsilon_{0.1})^{16/21} M_7^{2/21}~~\rm{~~if~} b=0,\label{e:gas/radb0}\\
r_3^{{\rm es}/{\rm ff}} &=& 4.10 \,\tilde\mu_0^{-1/3} \tilde f_{T}^{17/12} (\tilde \kappa_{\rm ff}/\tilde \kappa_{\rm es})^{-2/3} (\dot m_{0.1}/\epsilon_{0.1})^{2/3}\label{e:es/ff}.
\end{eqnarray}

Note that the middle and outer regions differ only in their opacity
laws, and the equations in these two regions are equivalent (this can
be seen by setting $\tilde \kappa_{\rm es}\equiv \tilde \kappa_{\rm
ff}\kappa_{\rm ff}(r)/\kappa_{\rm es}$). Since $\Sigma$, $H$, $\rho$,
and $T$ scale with a low power of $\tilde \kappa_{\rm ff}$, the radial
dependence ends up being similar in the middle and outer regions. The
distinction between these equations is nevertheless useful, since we
can assume that $\tilde \kappa_{\rm es}\rightarrow 1$ and $\tilde
\kappa_{\rm ff}\rightarrow 1$ are constants in the middle and outer
regions, respectively.

We emphasize that equations~(\ref{e:Sigma_inb1})-(\ref{e:H_out})
represent only a very non-exhaustive subset of solutions even for
radiatively efficient steady thin accretion disks.  In particular, at
large radii, there are several effects that can invalidate the disk
model described by these equations.  First, these solutions assume
that the self--gravity of the disk is negligible.  This assumption
becomes invalid at radii where the Toomre $Q$--parameter equals unity,
\begin{eqnarray}
r^{\rm sg}_3 &=& 12.6 \,\tilde\mu_0^{-8/9}\tilde\mu_e^{14/27} \tilde
f_T^{20/9} \tilde\kappa_{\rm es}^{2/9} \alpha_{0.3}^{8/9} \left(\dot
m_{0.1}/\eff_{0.1}\right)^{-8/27} M_7^{-26/27} \rm{~~~~if~} \tilde\kappa_{\rm es}\rightarrow1\\
r^{\rm sg}_3 &=& 30.99 \tilde\mu_0^{-1} \tilde\mu_e^{28/45} \tilde f_T^{143/60}
\tilde \kappa_{\rm ff}^{2/15} \alpha_{0.3}^{28/45} \left(\dot m_{0.1}/\eff_{0.1}\right)^{-22/45} M_7^{52/45}  \rm{~~~~if~} \tilde\kappa_{\rm ff}\rightarrow 1.
\end{eqnarray}

Beyond these radii, the disk is commonly believed to be unstable to
fragmentation.  Second, at large radii, the disk can also become
optically thin \cite[see][where solutions can be obtained by fixing
the Toomre parameter in the outermost region at $Q\equiv
1$]{sirkogoodman03}.  At these binary separations, the disks may not
actually be geometrically thin \citep{dot08,mke08}, and slim or thick
solutions might instead be relevant.  Third, beyond the radii where
the disk temperature falls below $\approx 10^4$K, the gas becomes
neutral. The corresponding change in opacity will modify the disk
structure, and the disk may become susceptible to ionization
instabilities (although see \citealt{mq01}).  Finally, at large radii
(where the orbital velocity $\gsim 100$ km/s), the gravitational
potential of the galaxy can no longer be ignored.  These regimes,
however, turn out to correspond to separations larger than we are
interested in the present paper, for BH masses above $\approx
10^5~{\rm M_\odot}$ (as will be shown in Figures~\ref{fig:tres_q1} and
\ref{fig:tres_q0.01} below).

\subsubsection{Comparison with Other Results}

We have verified our solutions numerically by substituting them back
into the fundamental conservation equations of thin accretion disks
\citep{shapiroteukolsky,accretionpower}.  Moreover,
equations~(\ref{e:Sigma_inb1})-(\ref{e:H_middle}) agree with those
quoted in \citet{goodmantan04}.\footnote{It appears that
\citet{goodmantan04} contain the following typographical errors:
$p\propto \alpha^{4/5} \mu^{-4/5} M_7^{4/5} r_3^{-18/5}$ in their
eq.~20 should be $p\propto \alpha^{-4/5} \mu^{4/5} M_7^{-4/5}
r_3^{-18/5}$, and $c_s\propto\kappa^{1/5} l_{\rm Edd} M_7^{0}
r_3^{-9/10}$ in their eq.~21 should be $c_s\propto \kappa l_{\rm Edd}
M_7^{0} r_3^{-3/2}$ and $c_s\propto \kappa^{1/10} l_{\rm Edd}^{1/5}
M_7^{-1/10} r_3^{-9/20}$ in the inner and middle regions,
respectively.}

Equations~(\ref{e:Sigma_inb0})-(\ref{e:H_in}) are also consistent with
\citet[page 441]{shapiroteukolsky}, for the $P=P_{\rm rad}$,
$\kappa=\kappa_{\rm es}$, $b=0$ model.  It is also reassuring that
equations~(\ref{e:Sigma_out})-(\ref{e:H_out}) are consistent with those
in \citet[Sec. 8.1, p. 244]{accretionpower}.\footnote{However, there appears to be a
typographical error in their quoted scaling $H\propto \dot
M_{26}^{3/10}$, which should instead read as $H\propto \dot
M_{26}^{3/20}$, so that $\Sigma=2\rho H$ is satisfied for all $\dot
M_{26}$.}  Also note that, owing to the weak dependence on
$\tilde\kappa_{\rm ff}$, our numerical factors are very similar to
those in \citet{accretionpower}, even though $\kappa_{\rm ff}$ is
defined to be two orders of magnitude larger there than the value we
adopted here (to be consistent with most other textbooks).

\subsection{Binary -- Disk Evolution}

Here we collect and summarize the most important formulae describing
the interaction between a binary and the accretion disk in order to
identify the mechanism that drives the orbital decay of the binary
during the final stages of the merger, as a function of binary
separation (the choices being GW driven inspiral and tidal--viscous
torques).  This will allow us to explicitly compute the residence time
$t_{\rm res}\equiv -R(dR/dt)^{-1}$ that an individual binary spends at
each orbital separation $R$, or at the corresponding orbital timescale
$t_{\rm orb}$.

The formulae collected in this section will also allow us to quantify
the binary separation at which the viscous evolution of the disk is
decoupled from the increasingly rapid, GW--driven orbital decay of the
binary. We provide results for $\alpha$ and $\beta$--disks, and give
analytic results as a function of binary and disk parameters.

\subsubsection{Disk- versus Secondary--Dominated Orbital Decay}
\label{subsec:diskorsecondary}

In general, the evolution of a SMBH binary in a thin disk is analogous
to planetary migration (see, e.g. \citealt{arm07}).  In the limit of a
very low--mass companion ($q\ll 1$), the interaction between the
planet and the disk is linear. In addition to co-rotation resonances,
the density waves excited in the gas at discrete Lindblad resonances
with the binary exert a large net torque on the binary, leading to
rapid, so--called Type--I migration, which occurs on a time--scale
much shorter than the local viscous time--scale (e.g.,
\citealt{tanaka02,arm07}).

If the binary is massive enough for the tidal torque to dominate over
the viscous torque in the disk, the interaction becomes non--linear,
and a gap is opened in the disk, extending to the outer radius
$R_{\lambda}=\lambda R$.  The condition for a gap to open is that the
mass ratio exceeds the critical value $q\gsim \max\{(H_{0}/R_{0})^3,
(10 \alpha)^{1/2} (H_{0}/R_{0})^{5/2}\}$ \citep[e.g.][note that $H_0$
is evaluated at the position of the secondary $R_0$]{rafikov02}.  For
binaries that are not in the GW--driven regime, and for which the disk
mass exceeds the mass of the secondary (see below), this typically
translates into the very modest requirement $q\gsim 10^{-7}$. This is
satisfied for all SMBH binaries that may produce the electromagnetic
signatures we discuss below.  The exceptions are the so--called
extreme mass--ratio binary inspirals (EMRI's) with $q\lsim 10^{-7}$
(i.e. a stellar--mass object coalescing with a SMBH). In this paper,
we focus on SMBH binaries, and therefore in the rest of this paper, we
neglect Type I migration.

If the secondary's mass satisfies the above gap--opening threshold,
but is still small compared to the local disk mass, then it acts as an
angular momentum bridge for the disk, and the secondary's orbital evolution 
is simply determined by the viscous diffusion time,
\begin{equation} \label{e:Armitage_Natarayan}
 t_{\nu} = -\frac{r}{\dot r_{\nu}}= \frac{2}{3}  \frac{R_0^2}{\nu_0}
 = 2\pi \frac{R_0^2 \Sigma_0}{\dot M}.
\end{equation}
where we have used $\dot M=3\pi \nu_0\Sigma_0$ which follows directly
from angular momentum conservation in steady disks \citep{ss73}.  The
orbital decay of the binary in this limit is analogous to {\it
disk--dominated Type-II planetary migration}.

In practice, the assumption that the local disk mass exceeds the
secondary's mass often fails. In this case, analogous to {\it
  ``planet--dominated'' Type-II migration}, the angular momentum of
the binary can still be absorbed by the gaseous disk outside the gap,
and the viscosity of the gas can drive the binary toward
merger. However, migration is slower, and the time--scale in this
regime, $t_{\rm s}$, is longer than $t_{\nu}$. An estimate of the
slowing factor is $q_B^{-k}$, where
\begin{equation}
\label{e:q_B}
q_B = \frac{4\pi R_0^2 \Sigma_0}{\mu} =\frac{2\dot M}{\mu}t_{\nu} =
\frac{8\dot m}{\q} \frac{t_{\nu}}{t_{\rm Edd}}
\end{equation}
is a measure of the lack of local disk--mass dominance \citep[but note
that our $q_B$ is denoted by ``$B$'' in their original
definition]{syerclarke}, which is less than unity in this case, and
$k$ is a constant defined as
\begin{equation}\label{e:k}
 k=\left\{
 \begin{array}{cc}
   1-\left(1+\frac{\partial \ln \Sigma}{\partial \ln \dot M}\right)^{-1}  &\rm{~~~~if~} q_B\leq 1 \\
   0 &\rm{~~~~if~} q_B> 1
 \end{array}\right\}
 =\left\{
 \begin{array}{cc}
   3/8  &\rm{~~~~if~} q_B\leq 1 ~\&~ \tilde\kappa_{\rm es}\rightarrow 1  \\
   7/17  &\rm{~~~~if~} q_B\leq 1 ~\&~ \tilde\kappa_{\rm ff}\rightarrow 1  \\
   0 &\rm{~~~~if~} q_B> 1
 \end{array}\right\}.
\end{equation}

Thus, the separation of the binary in this case is driven inward
on the timescale
\begin{equation}\label{e:t_II}
 t_{\rm s} = -\frac{r}{\dot r_{\rm s}} = {q_B}^{-k} t_{\nu}
 = \left(\frac{\q}{8\dot m}\frac{t_{\rm Edd}}{t_{\nu}} \right)^{k} t_{\nu}\rm{~~~~if~}q_B\leq 1.
\end{equation}

Note that the viscous time--scale $t_{\nu}$ in disk--dominated limit
(eq.~\ref{e:Armitage_Natarayan}) should be evaluated at the position
of the secondary $R_0=R/(1+q)$, while the quantities entering the
time--scale $t_{\rm s}$ for the secondary--dominated type-II migration
of more massive binaries ($q_B\leq 1$, eq.~\ref{e:t_II}), should be
evaluated at the outer edge of the cavity, $R_{\lambda}=\lambda R$
\citep{mm08}.  In order to avoid a discontinuous jump in the migration
time--scale at the $q_B=1$ transition, below we will omit this
distinction, and evaluate both time--scales at $R_{\lambda}=\lambda
R$.

\subsubsection{Evolution of Individual Binaries From Large to Small Radii}

Using the steady thin disk model outlined above, we can calculate the
rate at which the binary is driven inward by the gas. We will also
estimate the rate at which the inner edge of the punctured gaseous
disk follows the binary due to its viscosity.  From the preceding
discussion, we see that both the viscous time--scale and the orbital
decay rate depend on whether the binary is located in the
inner/middle/outer region of the disk; and also on whether the local
disk mass is larger/smaller than the mass of the smaller SMBH.  For
completeness, we here obtain and quote the residence time as a
function of orbital radius and orbital time, in each of these
$3\times2=6$ regimes.  We then construct the self--consistent
evolution of individual binaries, with different masses and
mass-ratios, across the relevant regimes.

We first consider the timescale $t_{\nu}$, and assume that the
secondary perturbs the disk at the radius $R_\lambda=\lambda R$.  This
is the relevant regime initially, at large binary separations, when
the disk mass enclosed within the secondary's orbit is large.  In this
regime, we find,
\begin{eqnarray}\label{e:t_nur-1}
t_{\nu} &=& (2.82 \times 10^{7}\yr)\times
   \tilde\kappa_{\rm es}^{-2} \tilde f_{T}^{-2} \alpha_{0.3}^{-1}
   \left(\frac{\dot m_{0.1}}{\epsilon_{0.1}}\right)^{-2} M_7 \lambda^{7/2}
   r_3^{7/2} \rm{~~if~} b=0 \rm{~and~} r_3 \lsim r_3^{{\rm
   gas}/{\rm rad}}\\
\nonumber
t_{\nu} &=& (5.96 \times 10^{4}\yr)\times
 \tilde\mu_e^{1/5} \tilde\mu_0^{4/5} \tilde\kappa_{\rm es}^{-1/5}
 \tilde f_{T}^{-2} \alpha_{0.3}^{-4/5} \left(\frac{\dot
 m_{0.1}}{\epsilon_{0.1}}\right)^{-2/5} M_7^{6/5} \lambda^{7/5}
 r_3^{7/5} \\
&& \left\{\begin{array}{l} \rm{~~if~} r_3^{{\rm gas}/{\rm
 rad}}\ll r_3 \lsim r_3^{{\rm es}/{\rm ff}},\\ \rm{~~or~if~} b=1
 \rm{~and~} r_3 \lsim r_3^{{\rm gas}/{\rm rad}}
\end{array}
\right.\\
\nonumber
t_{\nu} &=& (7.37 \times 10^{4}\yr)\times
\tilde\mu_e^{1/5}\tilde\mu_0^{3/4}\tilde\kappa_{\rm ff}^{-1/10} \tilde f_{T}^{-143/80}   \alpha_{0.3}^{-4/5} \left(\frac{\dot m_{0.1}}{\epsilon_{0.1}}\right)^{-3/10} M_7^{6/5} \lambda^{5/4} r_3^{5/4}
\\
&& \rm{~~~~if~} r_3 \gsim r_3^{{\rm es}/{\rm ff}}\label{e:t_nur-4}.
\end{eqnarray}
The above can be expressed as a function of the orbital time of the binary,
\begin{equation}\label{e:t_orb}
 t_{\rm orb} = \frac{2\pi}{\Omega} = \sqrt{8}\pi \frac{R_S}{\rm c} r^{3/2}  = 2.81 \times 10^5 \frac{R_S}{c} r_{3}^{3/2}
= (0.88\, M_7 r_{3}^{3/2})\, {\rm yr},
\end{equation}
which results in
\begin{eqnarray}\label{e:t_nuorb-1}
\nonumber
  t_{\nu} &=& (7.48 \times 10^{5}\yr)\times \tilde\kappa_{\rm es}^{-2}
 \tilde f_{T}^{-2} \alpha_{0.3}^{-1} \left(\frac{\dot
 m_{0.1}}{\epsilon_{0.1}}\right)^{-2} M_7^{-4/3} \lambda^{7/2}
 \left(\frac{t_{\rm orb}}{\yr}\right)^{7/3} \\
&& \rm{~~~~if~} b=0
 \rm{~and~} r_3 \lsim r_3^{{\rm gas}/{\rm rad}}\\
\nonumber
   t_{\nu} &=& (6.73 \times 10^{4}\yr)\times \tilde\mu_e^{1/5}
 \tilde\mu_0^{4/5} \tilde\kappa_{\rm es}^{-1/5} \tilde f_{T}^{-2}
 \alpha_{0.3}^{-4/5} \left(\frac{\dot
 m_{0.1}}{\epsilon_{0.1}}\right)^{-2/5} M_7^{4/15} \lambda^{7/5}
 \left(\frac{t_{\rm orb}}{\yr}\right)^{14/15} \\
&& \left\{\begin{array}{l}
 \rm{~~if~} r_3^{{\rm gas}/{\rm rad}}\ll r_3 \lsim r_3^{{\rm
 es}/{\rm ff}},\\ \rm{~~or~if~} b=1 \rm{~and~} r_3 \lsim r_3^{{\rm
 gas}/{\rm rad}}
\end{array}
\right.\\
\nonumber
   t_{\nu} &=& (8.21 \times 10^{4}\yr)\times
\tilde\mu_e^{1/5}\tilde\mu_0^{3/4}\tilde\kappa_{\rm ff}^{-1/10} \tilde
f_{T}^{-143/80} \alpha_{0.3}^{-4/5} \left(\frac{\dot
m_{0.1}}{\epsilon_{0.1}}\right)^{-3/10} M_7^{11/30} \lambda^{5/4}
\left(\frac{t_{\rm orb}}{\yr}\right)^{5/6}\\
&& \rm{~~~~if~} r_3 \gsim
r_3^{{\rm es}/{\rm ff}}\label{e:t_nuorb-4},
\end{eqnarray}

The measure of disk dominance can be calculated by substituting the viscous time--scale into equation~(\ref{e:q_B}),
\begin{eqnarray}\label{e:q_B-1}
\nonumber
q_B &=& (1.20 \times 10^{-3})\, \tilde\mu_e^{-1} \tilde\kappa_{\rm es}^{-2} \tilde
 f_{T}^{-2} \alpha_{0.3}^{-1} \left(\frac{\dot
 m_{0.1}}{\epsilon_{0.1}}\right)^{-1} M_7 \q^{-1} \lambda^{7/2}
 r_3^{7/2} \\
&& \rm{~~~~if~} b=0 \rm{~and~} r_3 \lsim r_3^{{\rm
 gas}/{\rm rad}}\\
\nonumber
q_B &=& 0.011\, \tilde\mu_e^{-4/5} \tilde\mu_0^{4/5}
 \tilde\kappa_{\rm es}^{-1/5} \tilde f_{T}^{-2} \alpha_{0.3}^{-4/5}
 \left(\frac{\dot m_{0.1}}{\epsilon_{0.1}}\right)^{3/5} M_7^{6/5}
 \q^{-1} \lambda^{7/5} r_3^{7/5} \\
&& \left\{\begin{array}{l} \rm{~~if~}
 r_3^{{\rm gas}/{\rm rad}}\ll r_3 \lsim r_3^{{\rm es}/{\rm ff}},\\
 \rm{~~or~if~} b=1 \rm{~and~} r_3 \lsim r_3^{{\rm gas}/{\rm rad}}
\end{array}
\right.\\
q_B &=& (1.49 \times
10^{-3})\,\tilde\mu_e^{-4/5}\tilde\mu_0^{3/4}\tilde\kappa_{\rm
ff}^{-1/10} \tilde f_{T}^{-143/80} \alpha_{0.3}^{-4/5}
\left(\frac{\dot m_{0.1}}{\epsilon_{0.1}}\right)^{7/10}
M_7^{6/5}\q^{-1} \lambda^{5/4} r_3^{5/4}\nonumber\\
 && \rm{~~~~if~} r_3 \gsim
r_3^{{\rm es}/{\rm ff}}\label{e:q_B-4}.
\end{eqnarray}
In order to decide whether the evolution indeed follows the
''disk--dominated'' decay (on the viscous timescale $t_{\nu}$) or
the secondary--dominated decay (on the longer time--scale
$t_{\rm s}$), one should examine whether $q_B> 1$ or $q_B\leq 1$ is
satisfied, respectively. From equations~(\ref{e:q_B-1})-(\ref{e:q_B-4}),
we find that the transition occurs at
\begin{eqnarray}\label{e:q_B-1b}
\nonumber
r_3^{{\nu}/{\rm s}} &=& 3.61\, \tilde\mu_e^{2/7} \tilde\kappa_{\rm
es}^{4/7} \tilde f_{T}^{4/7} \alpha_{0.3}^{2/7} \left(\frac{\dot
m_{0.1}}{\epsilon_{0.1}}\right)^{2/7} M_7^{-2/7} \q^{2/7} \lambda^{-1}\\
&& \rm{~~~~if~} b=0 \rm{~and~} r_3 \lsim r_3^{{\rm gas}/{\rm rad}}\\
\nonumber
r_3^{{\nu}/{\rm s}} &=& 121\, \tilde\mu_e^{4/7} \tilde\mu_0^{-4/7}
\tilde\kappa_{\rm es}^{1/7} \tilde f_{T}^{10/7} \alpha_{0.3}^{4/7}
\left(\frac{\dot m_{0.1}}{\epsilon_{0.1}}\right)^{-3/7} M_7^{-6/7}
\q^{5/7} \lambda^{-1} \\
&& \left\{\begin{array}{l} \rm{~~if~} r_3^{{\rm
gas}/{\rm rad}}\ll r_3 \lsim r_3^{{\rm es}/{\rm ff}},\\
\rm{~~or~if~} b=1 \rm{~and~} r_3 \lsim r_3^{{\rm gas}/{\rm rad}}
\end{array}
\right.\\
\nonumber
r_3^{{\nu}/{\rm s}} &=&
182\,\tilde\mu_e^{16/25}\tilde\mu_0^{-3/5}\tilde\kappa_{\rm ff}^{2/25}
\tilde f_{T}^{143/100} \alpha_{0.3}^{16/25} \left(\frac{\dot
m_{0.1}}{\epsilon_{0.1}}\right)^{-14/25} M_7^{-24/25} \q^{4/5}
\lambda^{-1} \\
&& \rm{~~~~if~} r_3 \gsim r_3^{{\rm es}/{\rm
ff}}\label{e:q_B-4b},
\end{eqnarray}
Note that with the exception of very unequal masses $q\lesssim 0.01$,
the transition takes place well in the outer region of the disk, with
$r_3^{{\nu}/{\rm s}}\gsim 10$.  At smaller radii, the binary is
driven viscously on the timescale $t_{\rm s}$ (rather than
$t_{\nu}$).

The ``secondary---dominated'' Type-II decay timescales relevant at
these radii can be obtained by substituting
equations~(\ref{e:t_nur-1})-(\ref{e:t_nur-4}) into
equation~(\ref{e:t_II})
\begin{eqnarray}\label{e:t_IIr-1}
\nonumber
t_{\rm s} &=& (6.15 \times 10^{6}\yr)\times
   \tilde\mu_e^{3/8}\tilde\kappa_{\rm es}^{-5/4} \tilde f_{T}^{-5/4}
   \alpha_{0.3}^{-5/8} \left(\frac{\dot
   m_{0.1}}{\epsilon_{0.1}}\right)^{-13/8} M_7^{5/8} \q^{3/8} \lambda^{35/16}
   r_3^{35/16} \\
&& \rm{~~~~if~} b=0 \rm{~and~} r_3 \lsim r_3^{{\rm
   gas}/{\rm rad}}\\
%
t_{\rm s} &=& (7.40 \times 10^{5}\yr)\times
 \tilde\mu_e^{1/2} \tilde\mu_0^{1/2} \tilde\kappa_{\rm es}^{-1/8}
 \tilde f_{T}^{-5/4} \alpha_{0.3}^{-1/2} \left(\frac{\dot
 m_{0.1}}{\epsilon_{0.1}}\right)^{-5/8} M_7^{3/4} \q^{3/8}
 \lambda^{7/8} r_3^{7/8} \nonumber \\
&& \left\{\begin{array}{l} \rm{~~if~}
 r_3^{{\rm gas}/{\rm rad}}\ll r_3 \lsim r_3^{{\rm es}/{\rm ff}},\\
 \rm{~~or~if~} b=1 \rm{~and~} r_3 \lsim r_3^{{\rm gas}/{\rm rad}}
\end{array}
\right.\\
%
t_{\rm s} &=& (1.07 \times 10^{6}\yr)\times
   \tilde\mu_e^{9/17}\tilde\mu_0^{15/34}\tilde\kappa_{\rm ff}^{-1/17}
   \tilde f_{T}^{-143/136} \alpha_{0.3}^{-8/17} \left(\frac{\dot
   m_{0.1}}{\epsilon_{0.1}}\right)^{-10/17} M_7^{12/17} \q^{7/17}
   \lambda^{25/34} r_3^{25/34} \nonumber \\
&& \rm{~~~~if~} r_3 \gsim r_3^{{\rm
   es}/{\rm ff}}\label{e:t_IIr-4},
\end{eqnarray}
or, in terms of $t_{\rm orb}$ using equation~(\ref{e:t_orb}),
\begin{eqnarray}\label{e:t_IIorb-1}
  t_{\rm s} &=& (3.60 \times 10^{6}\yr)\times
 \mu_e^{3/8}\tilde\kappa_{\rm es}^{-5/4} \tilde f_{T}^{-5/4}
 \alpha_{0.3}^{-5/8} \left(\frac{\dot
 m_{0.1}}{\epsilon_{0.1}}\right)^{-13/8} M_7^{-5/6} \q^{3/8} \lambda^{35/16}
 \left(\frac{t_{\rm orb}}{\yr}\right)^{35/24} \nonumber\\
&& \rm{~~~~if~} b=0
 \rm{~and~} r_3 \lsim r_3^{{\rm gas}/{\rm rad}}\\
   t_{\rm s} &=& (7.98 \times 10^{5}\yr)\times \tilde\mu_e^{1/2} \tilde\mu_0^{1/2} \tilde\kappa_{\rm es}^{-1/8} \tilde f_{T}^{-5/4} \alpha_{0.3}^{-1/2} \left(\frac{\dot m_{0.1}}{\epsilon_{0.1}}\right)^{-5/8}
 M_7^{1/6} \q^{3/8} \lambda^{7/8}  \left(\frac{t_{\rm orb}}{\yr}\right)^{7/12}\nonumber \\
&& \left\{\begin{array}{l}
   \rm{~~if~}  r_3^{{\rm gas}/{\rm rad}}\ll r_3 \lsim r_3^{{\rm es}/{\rm ff}},\\
 \rm{~~or~if~}  b=1 \rm{~and~} r_3 \lsim r_3^{{\rm gas}/{\rm rad}}
\end{array}
\right.\\
   t_{\rm s} &=& (1.14 \times 10^{6}\yr)\times \tilde\mu_e^{9/17}\tilde\mu_0^{15/34}\tilde\kappa_{\rm ff}^{-1/17} \tilde f_{T}^{-143/136}   \alpha_{0.3}^{-8/17} \left(\frac{\dot m_{0.1}}{\epsilon_{0.1}}\right)^{-10/17} M_7^{-11/51} \q^{7/17} \lambda^{25/34} \left(\frac{t_{\rm orb}}{\yr}\right)^{25/51}\nonumber\\
&&
\rm{~~~~if~} r_3 \gsim r_3^{{\rm es}/{\rm ff}}\label{e:t_IIorb-4},
\end{eqnarray}

Finally, at a still smaller radius, the orbital decay will be
dominated by gravitational wave emission. The GW--driven decay
timescale in the leading order (Newtonian) approximation, is
\begin{equation} \label{e:t_GW}
 t_{\rm GW} = -\frac{r}{\dot r_{\rm GW}}= \frac{5}{2} \frac{R_S}{\rm
 c} \q^{-1} r^4 = (1.11\times 10^7\yr) \times q_s^{-1} M_7^{-5/3}
 \left(\frac{t_{\rm orb}}{\yr}\right)^{8/3}.
\end{equation}
This approximation is adequate for our purposes, since post--Newtonian
corrections do not become appreciable until the final $\sim$ day of
the merger (see, e.g., Figure~5 in \citealt{paper2}).  Note that
$t_{\rm GW}$ defined above differs from the total time to merger,
(defined as the binary separation decreasing to zero), which is often
used in the literature, and which occurs at $t_{\rm GW}^{\rm
  merger}=t_{\rm GW}/4$.  What is the radius at which $t_{\rm GW}$
becomes smaller than the time--scale for Type-II orbital decay?  Let
us express this transition in terms of the radius $r_3$ that satisfies
$t_{\rm GW}=\beta_{{\rm GW}/{\rm s}} t_{\rm s}$, where $\beta_{{\rm
    GW}/{\rm s}}$ is a fixed constant of order unity:
\begin{eqnarray}\label{e:r_II/GW1}
\nonumber
  r_3^{{\rm s}/{\rm GW}} &=& 0.587\, \tilde\mu_e^{6/29}
  \tilde\kappa_{\rm es}^{-20/29} \tilde f_{T}^{-20/29} \alpha_{0.3}^{-10/29} \left(\frac{\dot m}{\epsilon_{0.1}}\right)^{-26/29} M_7^{-6/29} \q^{22/29}  \lambda^{35/29}
  \beta_{{\rm GW}/{\rm s}}^{16/29}\\
&&
  \rm{~~~~if~}  b=0 \rm{~and~} r_3 \lsim r_3^{{\rm gas}/{\rm rad}}\\
  r_3^{{\rm s}/{\rm GW}} &=& 0.470\, \tilde\mu_e^{4/25} \tilde\mu_0^{4/25} \tilde\kappa_{\rm es}^{-1/25} \tilde f_{T}^{-2/5} \alpha_{0.3}^{-4/25} \left(\frac{\dot m_{0.1}}{\epsilon_{0.1}}\right)^{-1/5}
M_7^{-2/25} \q^{11/25} \lambda^{7/25} \beta_{{\rm GW}/{\rm s}}^{8/25}\nonumber \\
&& \left\{\begin{array}{l}
   \rm{~~if~}  r_3^{{\rm gas}/{\rm rad}}\ll r_3 \lsim r_3^{{\rm es}/{\rm ff}},\\
 \rm{~~or~if~}  b=1 \rm{~and~} r_3 \lsim r_3^{{\rm gas}/{\rm rad}}
\end{array}
\right.\\
  r_3^{{\rm s}/{\rm GW}} &=& 0.545\, \tilde\mu_e^{6/37}\tilde\mu_0^{5/37}\tilde\kappa_{\rm ff}^{-2/111} \tilde f_{T}^{-143/444}   \alpha_{0.3}^{-16/111} \left(\frac{\dot m_{0.1}}{\epsilon_{0.1}}\right)^{-20/111} M_7^{-10/111} \q^{16/37} \lambda^{25/111} \beta_{{\rm GW}/{\rm s}}^{34/111}\nonumber \\
&&
   \rm{~~~~if~} r_3 \gsim r_3^{{\rm es}/{\rm ff}}\label{e:r_II/GW4}.
\end{eqnarray}
The corresponding critical radius is around $\sim 500 R_S$ for system
parameters near the assumed fiducial values.  The critical radius,
however, is significantly closer in for very massive, and very
unequal--mass binaries (i.e. for $M=10^9~{\rm M_\odot}$ and $q=0.01$;
see Fig.~\ref{fig:tres_r_q0.01} below).  Interestingly, the critical
radius is quite insensitive to the BH mass and accretion rate (i.e. to
$M_7$ and $\dot m$).  Note that the viscous timescale, $t_{\nu}$,
describing gas accretion, is faster than $t_{\rm s}$, which indicates
that at the time when GW starts driving the evolution, the viscous
inward diffusion of gas can initially still follow the binary.
However, the comparison of
equations~(\ref{e:t_nur-1})-(\ref{e:t_nur-4}) and
equation~(\ref{e:t_GW}) shows that as the binary orbit shrinks
further, the viscous time-scale always decreases less rapidly than the
GW inspiral timescale, so that eventually the evolution of the gaseous
disk will decouple from that of the binary.  Let us find the critical
radius, $r_3^{{\rm \nu}/{\rm GW}}$, where GW inspiral outpaces viscous
gas accretion.  We find that in most cases, this critical radius is
not relevant for the orbital decay of the BHs themselves, because the
transition to secondary--driven orbital decay always takes place
before GWs start dominating the decay.\footnote{The exceptions to this
  are the most--massive, $M>10^{10}~{\rm M_\odot}$, equal--mass
  binaries, and only if $b=1$ is assumed -- in this case, the
  GW--inspiral takes over in a radiation--pressure dominated disk, in
  the disk--dominated regime, i.e. before the transition to the
  secondary--dominated regime.}  However, this critical radius is
relevant for the behavior of the disk: it provides an estimate for the
time when the punctured disk decouples from the GW--driven binary, and
effectively stops evolving (and also for the size of the inner gap at
this time and onward).  By requiring $t_{\rm GW} = \beta_{{\rm
    GW}/{\nu}} t_{\nu}$, where $\beta_{{\rm GW}/{\nu}}$ is a constant
coefficient of order unity, we obtain:
\begin{eqnarray}\label{e:r_nu/GW1}
  r_3^{{\nu}/{\rm GW}} &=& 0.202\, \tilde\kappa_{\rm es}^{-4} \tilde f_{T}^{-4} \alpha_{0.3}^{-2} \left(\frac{\dot m}{\epsilon_{0.1}}\right)^{-4} \q^2  \lambda^{7} \beta_{{\rm GW}/{\nu}}^{2}
  \rm{~~~~if~}  b=0 \rm{~and~} r_3 \lsim r_3^{{\rm gas}/{\rm rad}}\\
r_3^{{\nu}/{\rm GW}} &=& 0.222\, \tilde\mu_e^{1/13} \tilde\mu_0^{4/13} \tilde\kappa_{\rm es}^{-1/13} \tilde f_{T}^{-10/13} \alpha_{0.3}^{-4/13} \left(\frac{\dot m_{0.1}}{\epsilon_{0.1}}\right)^{-2/13}
M_7^{1/13} \q^{5/13} \lambda^{7/13} \beta_{{\rm GW}/{\nu}}^{5/13}\nonumber \\
&&
\left\{\begin{array}{l}
   \rm{~~if~}  r_3^{{\rm gas}/{\rm rad}}\ll r_3 \lsim r_3^{{\rm es}/{\rm ff}},\\
 \rm{~~or~if~}  b=1 \rm{~and~} r_3 \lsim r_3^{{\rm gas}/{\rm rad}}
\end{array}
\right.\\
  r_3^{{\nu}/{\rm GW}} &=& 0.183\, \tilde\mu_e^{4/55}\tilde\mu_0^{3/11}\tilde\kappa_{\rm ff}^{-2/55} f_{T}^{-13/20}   \alpha_{0.3}^{-16/55} \left(\frac{\dot m_{0.1}}{\epsilon_{0.1}}\right)^{-6/55} M_7^{4/55} \tilde\q^{4/11} \lambda^{5/11} \beta_{{\rm GW}/{\nu}}^{4/11} \nonumber \\
&&
   \rm{~~~~if~} r_3 \gsim r_3^{{\rm es}/{\rm ff}}\label{e:r_nu/GW4}.
\end{eqnarray}

The appropriate choices for $\beta_{{\rm GW}/{\rm s}}$ and
$\beta_{{\rm GW}/{\nu}}$ are poorly known, but $\beta_{{\rm GW}/{\rm
    s}}$ may be reasonably taken to be $\sim 1$ when the binary is
first driven by GW emission, rather than by tidal interaction with the
gas.  The simplest choice for $\beta_{{\rm GW}/{\nu}}$, adopted in
many previous studies, is also $\beta_{{\rm GW}/{\nu}}=1$
\citep[e.g.][]{an05,loeb07}.  However, the gas inflow rate across the
edge of the central gap will be increased due to the steep density and
pressure gradient \citep{lp74}, which will delay the decoupling. This
motivated \citet{mp05} to adopt $\beta_{{\rm GW}/{\nu}}\sim 0.1$ (the
value describing the limiting case of an infinitely sharp edge).

Adopting $\beta_{{\rm GW}/{\rm s}}=1$ in
equations~(\ref{e:r_II/GW1})-(\ref{e:r_II/GW4}) then yields the radius
where the binary evolution changes from being viscosity--driven to
GW--driven, and $\beta_{{\rm GW}/{\nu}}= 0.1$ in
equations~(\ref{e:r_nu/GW1})-(\ref{e:r_nu/GW4}) gives the separation
at which the disk totally decouples from the binary and the radius of
the gap ``freezes''.  These expressions generalize the results of
\citet{mp05}, who restricted their analysis to the $b=1$ case, and
focused on the behavior of the disk at decoupling, rather than the
orbital evolution of the binary.  In particular, \citet{mp05} evaluate
disk conditions at the single radius at the edge of the gap, at the
time of decoupling, and do not discuss the transition from the disk--
to the secondary--dominated decay, or other details of the binary's
orbital decay.  The binary separation at decoupling is of order
$r_3^{{\rm visc}/{\rm GW}}\sim 0.1$ for both the gas pressure
dominated models and the radiation pressure dominated case with
$b=1$. In these cases, the transition between viscosity and GW--driven
decay and the disk decoupling take place in relatively quick
succession, since $r_3^{{\rm visc}/{\rm GW}}$ depends weakly on
$\beta_{{\rm GW}/{\nu}}$.  The delay between these two events is much
longer for the radiation pressure dominated regime when $b=0$, since
in this case the viscosity, which is proportional to the total, rather
than just the gas pressure, is much larger, and the gas can follow the
binary nearly all the way to merger (at least for large $\dot m$). In
this case, the result is also extremely sensitive to the accretion
rate and the binary mass ratio. Generically, for a fixed total binary
mass, the decoupling occurs at the largest separations for nearly
equal masses.

\begin{figure}[tbh]
\centering
\mbox{\includegraphics[width=12.5cm]{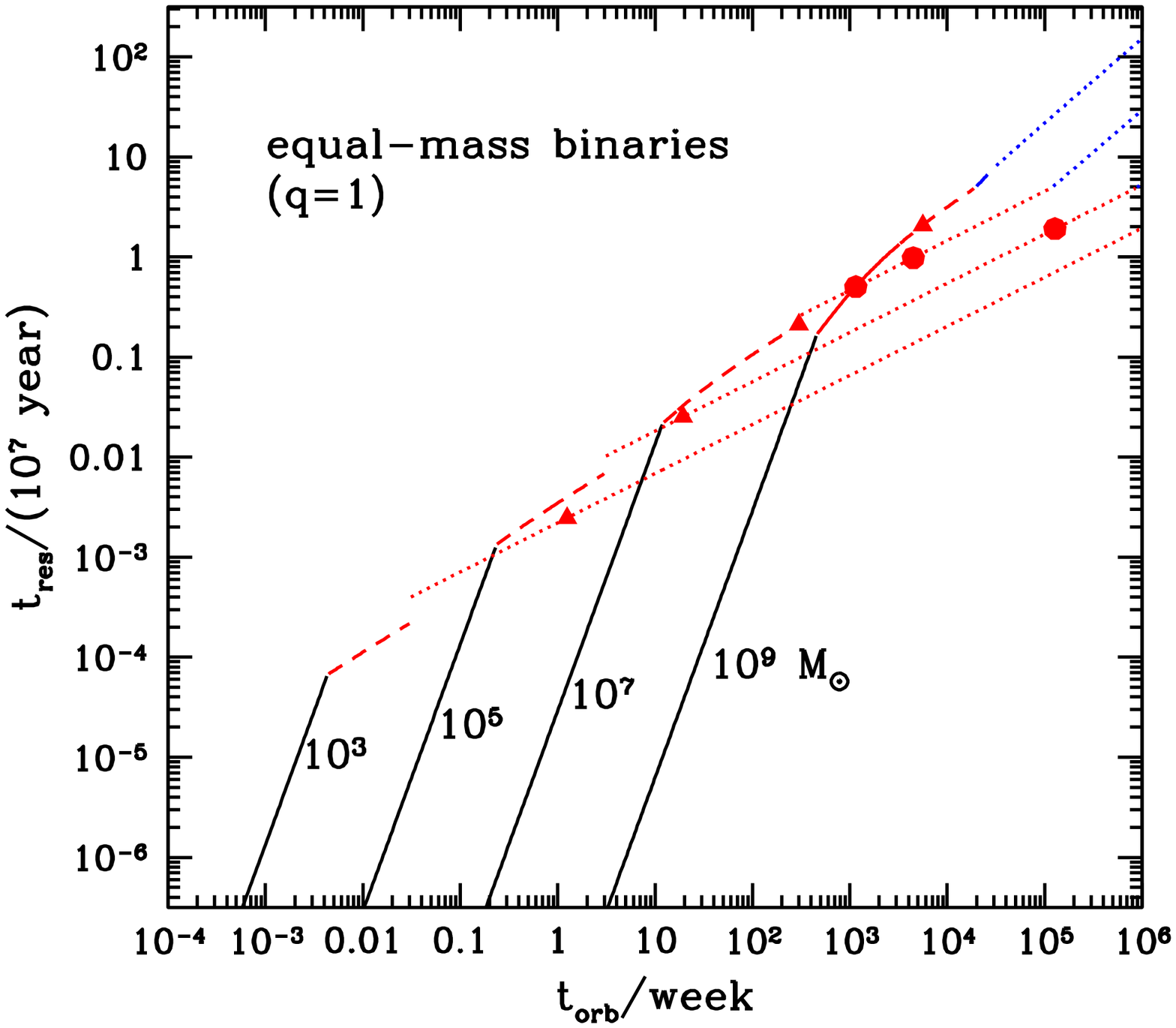}}
\caption{\label{fig:tres_q1} The evolution of equal--mass ($q=1$) SMBH
  binaries, embedded in a steady circumbinary disk, from large to
  small orbital separations. The figure shows the residence time
  $t_{\rm res}=-R/(dR/dt)$ that each binary spends at the radius where
  the orbital time is $t_{\rm orb}$.  The four curves correspond to
  binaries with total masses of $M=10^3, 10^5, 10^7$, and $10^9~{\rm
    M_\odot}$, as labeled. The large dots denote the critical radius
  beyond which the assumed circumbinary disk is unstable to
  fragmentation (Toomre parameter $Q<1$). Similarly, triangles denote
  radii beyond which the disk may be susceptible to ionization
  instabilities (the gas temperature falls below $10^4$K).  In each
  case, blue/red colors indicate whether the disk mass enclosed within
  the binary's orbit is larger/smaller than that mass of the
  secondary.  The dotted/dashed/solid portion of each curve indicates
  the outer/middle/inner disk region, respectively (as defined in
  \S~\ref{subsec:thindisks}).  For a binary located at redshift $z$,
  the redshifted values of $t_{\rm res}$ and $t_{\rm orb}$ (as
  measured on Earth), should be multiplied by a factor of $(1+z)$.  }
\end{figure}

\begin{figure}[tbh]
\centering
\mbox{\includegraphics[width=12.5cm]{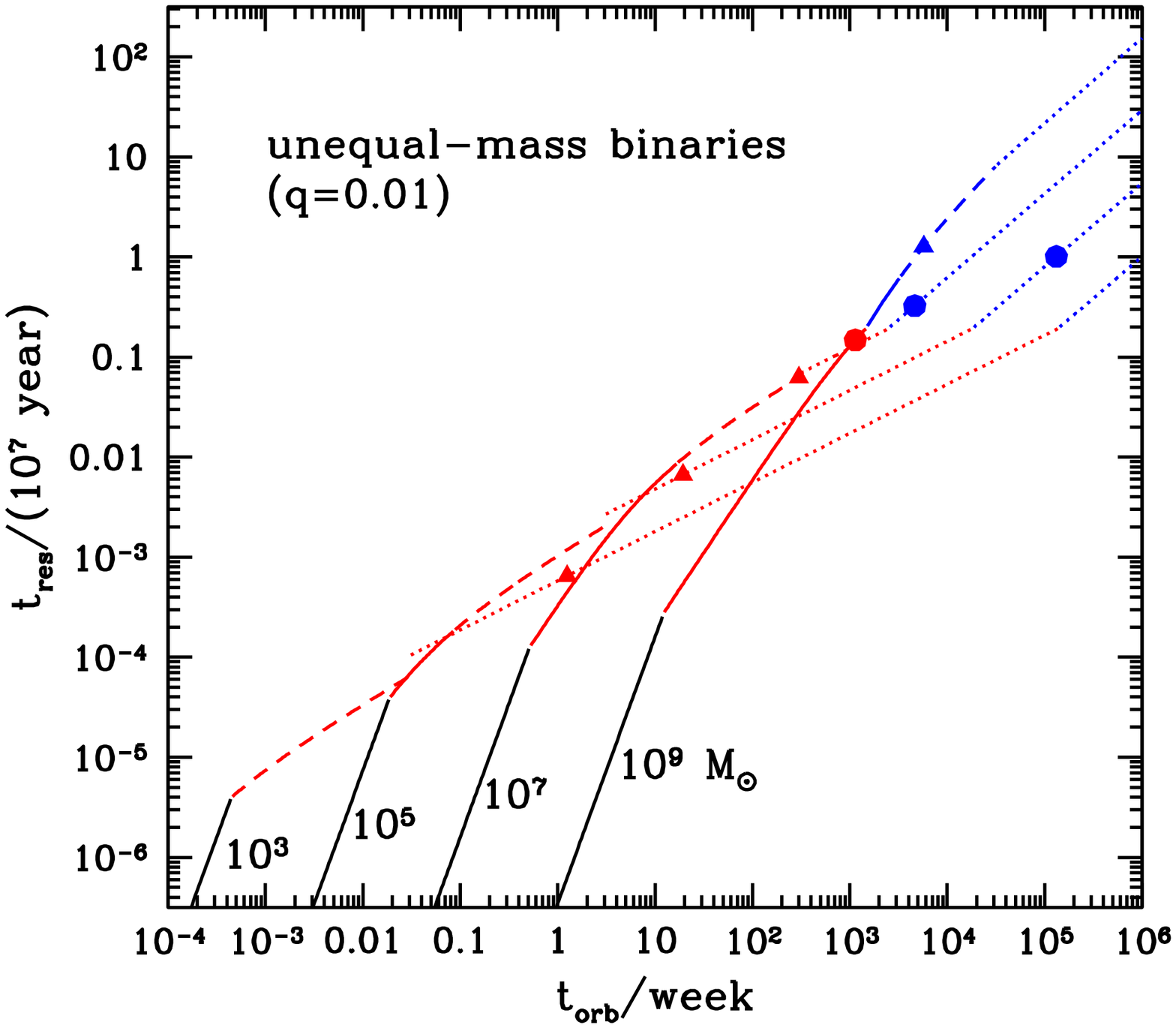}}
\caption{\label{fig:tres_q0.01} The figure shows the residence time
  $t_{\rm res}$ as in Figure~\ref{fig:tres_q1}, but for unequal--mass
  binaries ($q=0.01$).  }
\end{figure}

Interestingly, the decay rate of a given individual binary can
decelerate and accelerate during its evolution, according to the
variations in the local disk environment at each instantaneous binary
separation.  The evolutionary tracks of binaries with four different
choices for the total mass ($M=10^3, 10^5, 10^7$, and $10^9~{\rm
  M_\odot}$) and two different mass ratios ($q=1$ and $q=0.01$) are
shown in Figures~\ref{fig:tres_q1} and ~\ref{fig:tres_q0.01}.  In both
figures, we assume that the viscosity is proportional to the total
pressure ($b=0$).  The motivation for this choice is to illustrate the
effect of the additional radiation pressure--related viscosity on the
orbital decay (which is not present in the $b=1$ case).  We note that
a phenomenological $b=0$ disk is known to suffer from a formal thermal
instability \citep[e.g.][]{le74}; recent magnetohydrodynamical
simulations, however, found such disks thermally stable \citep[while
accounting for the correlation between viscosity and radiation
pressure][]{hirose09}.  These figures show the residence time as a
function of the orbital time.  They demonstrate that the evolution of
the binary in most cases proceeds through the following distinct
stages.

{\em (i) Disk--dominated viscous evolution.}
Initially, at large separations (shown in blue curves), the binary is
strongly coupled to the circumbinary disk and evolves on the viscous
time--scale $t_{\rm visc}$ (analogous to ``disk-dominated'' planetary
migration).  The radius of the gap follows the binary.  During this
stage, $t_{\rm res}\approx t_{\nu}$ is proportional to
$r^{7/5}$--$r^{7/2}$ (the range corresponding to the choice $b=0$ {\it
  vs.}  $b=1$) for radiation pressure, or $r^{7/5}$--$r^{5/4}$ (the
range corresponding to the choice of dominant opacity being electron
scattering or free--free absorption) for gas pressure dominated disks
(see eqs.~\ref{e:t_nur-1}--\ref{e:t_nur-4}).  These decay rates
translate into $t_{\nu}\propto t_{\rm orb}^{14/15}$--$t_{\rm
  orb}^{7/3}$, or $t_{\rm orb}^{14/15}$--$t_{\rm orb}^{5/6}$, in the
two cases respectively (see
eqs.~\ref{e:t_nuorb-1}--\ref{e:t_nuorb-4}).  Note, however, that for
nearly equal--mass binaries, the separations have to be quite large to
correspond to this disk--dominated regime -- falling into the outer
regions of the disk, which are unstable to fragmentation (the orbital
radii where the disks are marginally Toomre--stable are marked with
large dots).  Therefore, depending on the behavior of the gas disk
beyond this radius, this early stage of disk--dominated viscous
evolution may exist only for unequal--mass binaries. As shown in
Figure~\ref{fig:tres_q0.01}, disk--dominated viscous evolution may be
realized in a stable disk for binaries with $M\lsim 10^7~{\rm
  M_\odot}$ and $q\sim 0.01$; in these cases, the binaries are in the
free--free opacity and gas--pressure dominated regions of the disk, so
the relevant scaling is $t_{\nu}\propto r^{5/4} \propto t_{\rm
  orb}^{5/6}$.

{\em (ii) Secondary--dominated viscous evolution.}
As the binary separation shrinks below $R^{\nu/\rm s}\sim 10^5 R_S$
($10^3 R_S$) for mass ratios $q\sim 1$ ($q\sim 0.01$), the binary mass
starts to dominate over the local disk mass, and the binary evolves
more slowly, according to ``secondary--dominated'' decay (analogous to
``planet-dominated'' Type-II migration).  During this stage, the GW
emission is still negligible, and the decay time--scale can be
obtained from equations~(\ref{e:t_IIr-1})-(\ref{e:t_IIr-4}), and
$t_{\rm s}\propto r^{7/8}$--$r^{35/16}$ for radiation pressure (with
$b=0-1$), or $r^{7/8}$--$r^{25/34}$ for gas pressure dominated (with
electron scattering vs. free--free opacity) disks, implying that
$t_{\rm s}\propto t_{\rm orb}^{7/12}$--$t_{\rm orb}^{35/24}$, and
$t_{\rm orb}^{7/12}$--$t_{\rm orb}^{25/51}$, in the two cases
respectively (see eqs.~\ref{e:t_IIorb-1}--\ref{e:t_IIorb-4}).  As can
be seen from Figures~\ref{fig:tres_q1} and \ref{fig:tres_q0.01}, on
orbital time--scales between weeks to years, each of these scalings is
relevant for some choice of binary masses.  However, the transition to
GW--domination always takes place either in the ``inner'' or
``middle'' disk region.

{\em (iii) GW--dominated evolution.}
Still later, within the radius $R^{{\rm s}/{\rm GW}}\sim 500 R_S$ for
systems with parameters close to the fiducial values, the binary's
orbital evolution starts to be driven primarily by GWs, but the outer
edge of the gap can still diffuse inward and follow the binary.
During this stage, the decay time--scale is $t_{\rm GW}\propto
r^4\propto t_{\rm orb}^{8/3}$.

{\em (iv) Gas disk decoupled.}
Finally, within $R_S^{{\rm \nu}/{\rm GW}}\sim 100 R_S$ the binary is
entirely driven by GWs and the binary falls in much more quickly than
the outer edge of the gap is able to move inward.

The above ordering of events is valid for a broad range of binary and
disk parameters. Note that the ultimate fate of the gas inside the
binary's orbit is left unspecified in our considerations \citep[see,
e.g.][for a possible outcome]{an02}.

In addition to the above sequence of events describing the evolution
of individual binaries, several interesting conclusions can be drawn
from Figures~\ref{fig:tres_q1} and \ref{fig:tres_q0.01}.

\begin{enumerate}

\item {\em Coalescing binaries have a non-negligible abundance.}
  First, binaries with masses in the range $10^5-10^9 {\rm M_\odot}$
  may be both bright and common enough to be detectable in a survey,
  provided they have bright emission.  Indeed,
  Figures~\ref{fig:tres_q1} and \ref{fig:tres_q0.01} show that these
  binaries spend a non--negligible fraction ($\gsim 10^{-3}$) of their
  total fiducial lifetime of $10^7$ years at orbital time--scales
  between 1 day $\lsim t_{\rm orb}\lsim$ 1 year (the total lifetime
  will be justified below).  It is feasible, in principle, to look for
  variability on these time--scales, and the residence times shown on
  the figures suggest that these variables may not be uncommon among
  bright AGN.  We will discuss this possibility further in
  \S~\ref{sec:periodic} below.

\item {\em Disk-- and GW--driven evolution may both be observationally
    relevant.}  Figures~\ref{fig:tres_q1} and \ref{fig:tres_q0.01}
  also show that the transition from gas-- to GW--driven evolution can
  occur within this ``observational window''.  For example, at the
  fixed orbital time--scale of $t_{\rm orb}=10$ weeks, equal--mass
  binaries above $10^7 {\rm M_\odot}$ are GW--driven, and below this
  mass, they are gas--driven.

\item {\em Secondary--dominated evolution cannot be ignored.}
  Essentially all binaries at the orbital times relevant for actual
  surveys (again, between 1 day $\lsim t_{\rm orb}\lsim$ 1 year) that
  are gas--driven are in the regime of ``secondary--dominated''
  type-II orbital decay (referred to as stage {\it (ii)} above).
  Likewise, the transition from ``gas--driven'' to ``GW--driven''
  evolution always occurs from the ``secondary--dominated'' type-II
  decay regime. In previous works whose primary focus was on the
  behavior of gas at (and after) the time of decoupling
  (e.g. \citealt{mp05,loeb07}), this intermediary step, which is
  important for the orbital decay of the binary, is not discussed.

\item {\em Observed binaries could probe all three disk regions.}
  Interestingly, among the $10^5-10^9 {\rm M_\odot}$ binaries with 1
  day $\lsim t_{\rm orb}\lsim$ 1 year, it appears that all three of
  the disk regions (inner/middle/outer) enumerated in
  \S~\ref{subsec:thindisks} can be observationally relevant (i.e.,
  gas--driven binaries can be found in each of these three disk
  regions).

\item {\em Viscous evolution is non--negligible even in the LISA
    regime.}  The comparison of Figures~\ref{fig:tres_q1} and
  \ref{fig:tres_q0.01} shows that unequal--mass binaries evolve more
  rapidly when they are gas--driven. Consequently, they make the
  transition to the GW--driven stage quite late in their evolution.
  In particular, binaries enter {\it LISA}'s detection range at the
  approximate observed GW frequency of $f_{\rm GW}=0.03$ mHz.  This
  corresponds to an observed orbital time (on Earth) of $t_{\rm orb} =
  2/f_{\rm GW} = 0.11$ week.  We find that at this orbital time,
  viscous evolution is not necessarily
  negligible. Figures~\ref{fig:tres_q1} and \ref{fig:tres_q0.01} show
  that equal--mass binaries with $M \lsim 10^5 \,{\rm M_\odot}$, and
  $q=0.01$ binaries with $M \lsim 10^6 \,{\rm M_\odot}$ are just
  making the transition to the GW-driven regime as they enter the {\it
    LISA} band.

\item {\em Total decay time in a stable disk is consistent with quasar
    lifetime.}  As Figures~\ref{fig:tres_q1} and \ref{fig:tres_q0.01}
  show, the residence time at the radius at which $Q=1$ is, in all
  cases, close to (although somewhat shorter) than the fiducial quasar
  lifetime of $10^7$ years.  It is plausible that SMBHs become
  luminous, and act as quasars, only once they are embedded in stable
  circumbinary accretion disk.  The fact that it takes $\sim 10^7$
  years for the binary to evolve from the outer edge of a stable disk
  to coalescence is therefore consistent with the idea proposed in
  this paper, that there is a one--to--one correspondence between
  coalescing SMBHs and quasars (although, as mentioned above, there
  are caveats that can invalidate the steady disk models at the
  relevant large radii).

\end{enumerate}

\begin{figure}[tbh]
\centering
\mbox{\includegraphics[width=12.5cm]{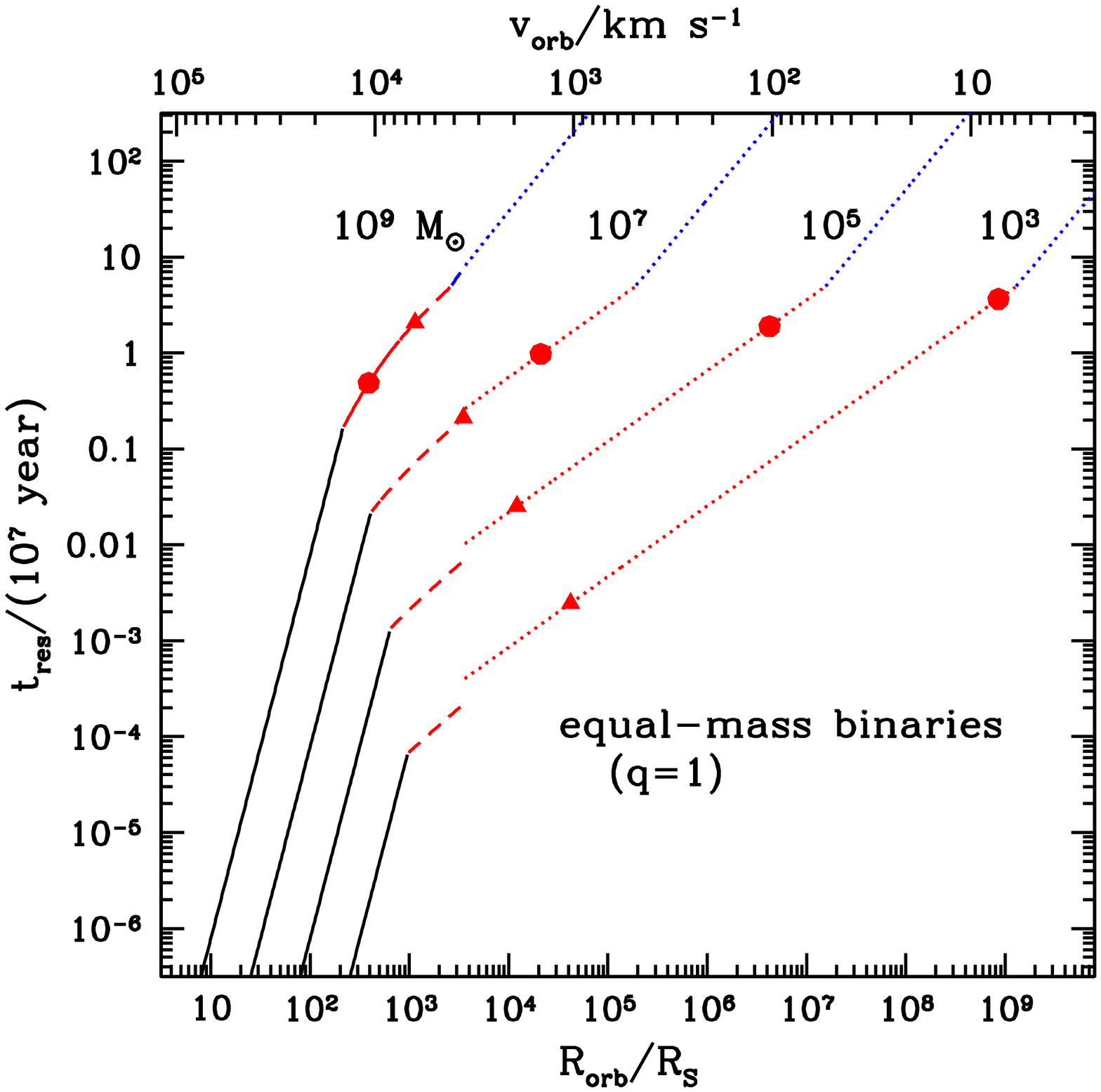}}
\caption{\label{fig:tres_r_q1} The residence time for equal--mass
binaries, as in Figure~\ref{fig:tres_r_q1}, except that $t_{\rm res}$
is here shown as a function of orbital separation $R$, in units of the
Schwarzschild radius $R_S$.  For reference, the $x$ axis labels on the
top show the orbital velocity corresponding to each value of $R/R_S$.}
\end{figure}

\begin{figure}[tbh]
\centering
\mbox{\includegraphics[width=12.5cm]{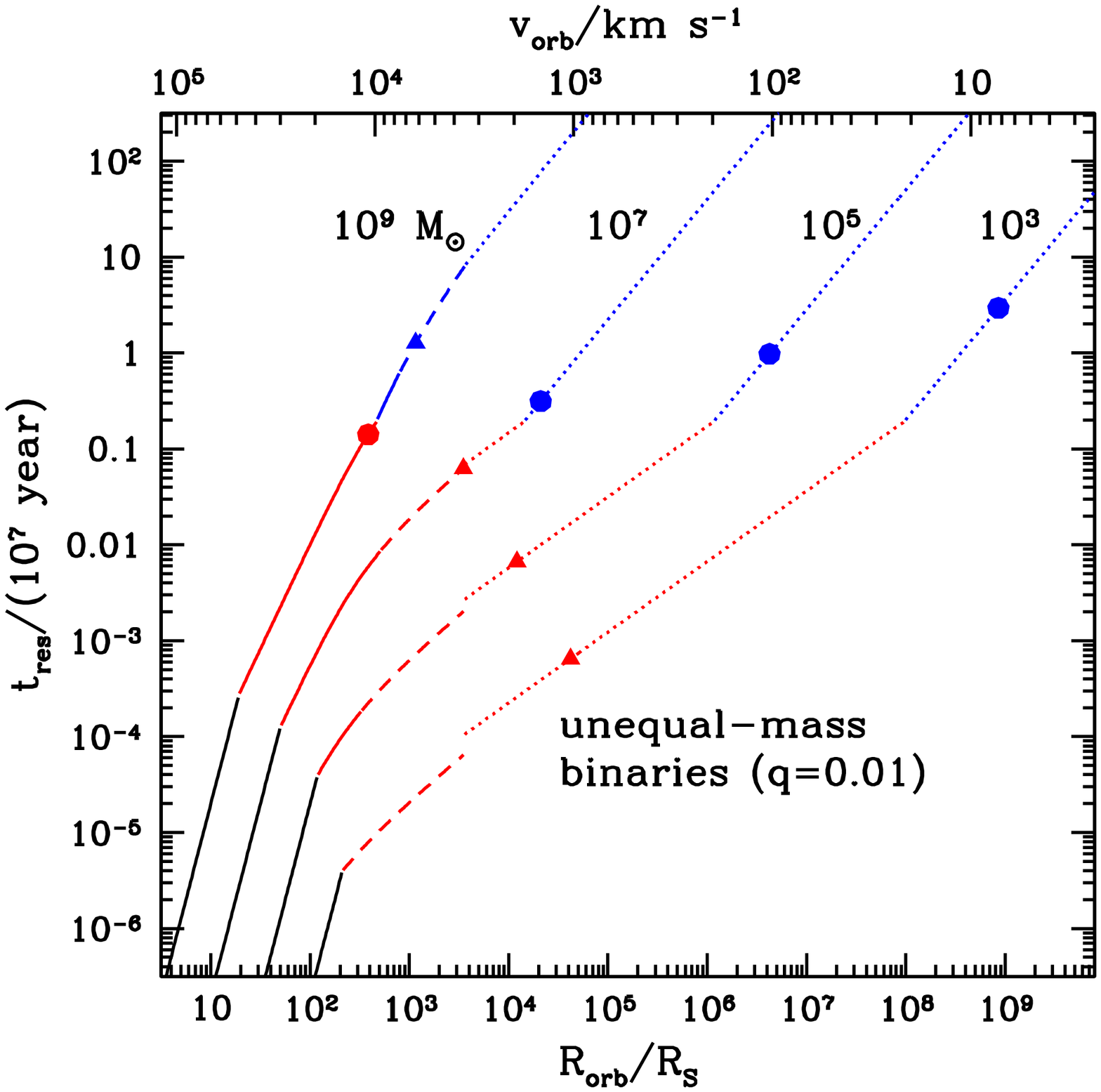}}
\caption{\label{fig:tres_r_q0.01} The residence time $t_{\rm res}$ as in
  Figure~\ref{fig:tres_r_q1}, except for unequal--mass binaries
  ($q=0.01$).  }
\end{figure}

The possible implication of conclusion no. 5 above for {\it LISA}
merits some further elaboration.  As discussed, e.g., in
\citet{sesana04}, individual binaries can contribute to the {\it LISA}
data stream in several ways.  Sources can be divided into two types,
based on whether they evolve significantly on a time--scale of $\sim
3$ years, the duration of the {\it LISA} experiment.  Binaries caught
at an orbital separation with short enough residence times for the
frequency--evolution to be measurable are sometimes referred to as
``gravitational sirens'' or ``gravitational inspirals''.
Figures~\ref{fig:tres_q1} and \ref{fig:tres_q0.01} show that during
the last several years of the coalescence, the orbital evolution is
always strongly GW--dominated, even for the lowest--mass BHs, and
therefore the GW waveform of these rapidly evolving sources (including
those whose actual coalescence is detected by {\it LISA}) will not be
affected by the gas disk.

Binaries that have a much longer residence time at some fixed
frequency in {\it LISA}'s band represent ``stationary'' sources whose
frequency remains roughly constant during the {\it LISA} mission
lifetime.  These sources could, in principle, be individually
detectable by {\it LISA}. However, in practice, they are likely to
accumulate sufficient signal--to--noise for detection only in the last
few hundred years of their coalescence (see, e.g., Figure 2 in
\citealt{sesana05} for the detectability of $q=0.1$ binaries as a
function of their look--back time from the merger).
Figures~\ref{fig:tres_q1} and \ref{fig:tres_q0.01} show that viscous
processes can significantly speed up the evolution of binaries only at
somewhat larger look--back times (note that the look--back time is 4
times shorter than the residence time in the pure GW--driven case).
The cumulative signal from a collection of faint stationary sources
can, however, still add up to an unresolved background that is
detectable, depending on the the cosmic evolution of the BH merger
rate and the instrumental noise of {\it LISA}.  The presence of the
gas disks could reduce any such background that is present (compared
to a prediction that assumes pure GW--driven evolution at {\it
  LISA}--frequencies).

In Figures~\ref{fig:tres_q1} and \ref{fig:tres_q0.01}, we have showed
the evolution of the binary as a function of its orbital period.  This
will be particularly useful for assessing the detectability of such
binaries in a survey for periodically variable sources
(\S~\ref{sec:periodic} below).  In Figures~\ref{fig:tres_r_q1} and
\ref{fig:tres_r_q0.01}, we show, instead, the evolution of the same
set of binaries, but as a function of their orbital separation.  The
$x$--axis on these figures is shown in units of $R_S$, with the
corresponding orbital velocities shown by the labels on the top axis.
This figure directly reveals that relatively more massive binaries
($M\gsim 10^7~{\rm M_\odot}$) spend a significant time at orbital
velocities of several thousand ${\rm km~s^{-1}}$.  Such orbital speeds
may be detectable in the spectra of individual sources, providing an
alternative to the detection based on periodic flux variations (see
\S~\ref{sec:alternatives} below).

\begin{figure}[tbh]
\centering
\mbox{\includegraphics[width=12.5cm]{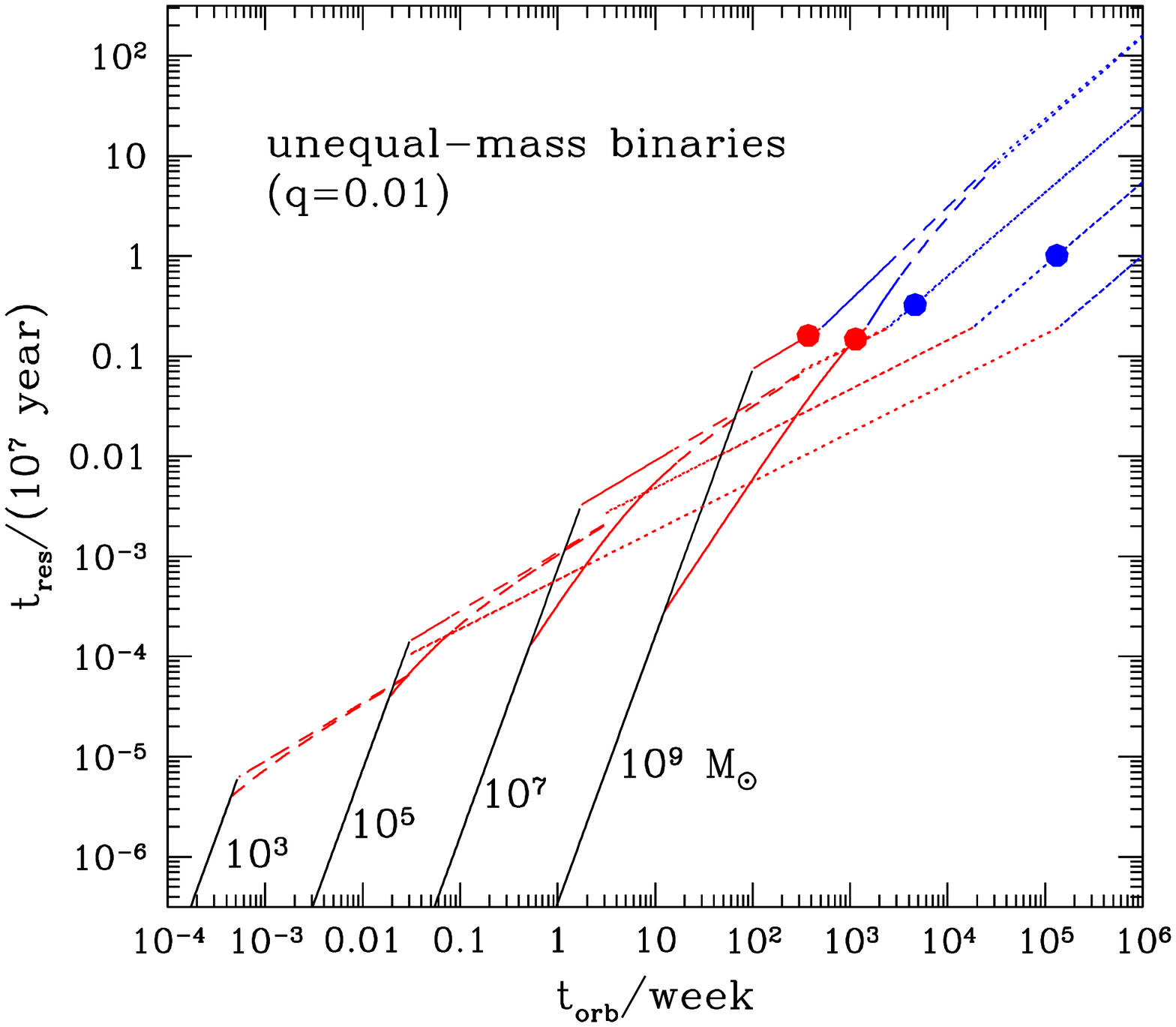}}
\caption{\label{fig:tres_b_q0.01} The residence time for equal--mass
binaries, as in Figure~\ref{fig:tres_r_q0.01}, except that for each
binary, we also show the results for $b=1$ (top curves), in addition
to the $b=0$ case (bottom curves, reproduced from
Fig.~\ref{fig:tres_r_q1}).  In the $b=1$ case, the viscosity is
proportional to the gas pressure, rather than the total pressure.
Once the binary approaches the radiation--pressure dominated regime
(shown in solid), the evolution is therefore slower than in the $b=0$
case.}
\end{figure}

Finally, the conclusions enumerated above also highlight the large
uncertainty in the residence times predicted in
Figures~\ref{fig:tres_q1} and \ref{fig:tres_q0.01}, caused by our
idealized treatment of ``secondary--dominated'' type-II orbital decay.
One immediate additional source of uncertainty is the choice of $b=0$
vs. $b=1$.  Before entering the GW--driven regime, most of the
equal--mass binaries (Fig.~\ref{fig:tres_q1}) are in the gas--pressure
dominated region of the disk, but unequal--mass binaries
(Fig.~\ref{fig:tres_q0.01}) are in the radiation--pressure dominated
region.  Therefore, whether the viscosity is proportional to the total
pressure or just the gas pressure makes little difference to the
near--equal mass binaries.  However, it makes a significant difference
for unequal--mass binaries with $M\gsim 10^6~{\rm M_\odot}$.  To show
this explicitly, in Figure~\ref{fig:tres_b_q0.01}, the upper vs. lower
curves contrast the evolution in the $b=1$ vs $b=0$ case,
respectively.  As expected, once the binary approaches the
radiation--pressure dominated regime, the evolution is significantly
slower in the $b=1$ case. The difference is most pronounced for the
most massive ($10^9~{\rm M_\odot}$) binary. For this system, the
transition to GW--domination also occurs at a larger orbital time
($\approx 10^2$ weeks for $b=1$, vs. $\approx 10$ weeks for $b=0$).

\subsection{Type II Orbital Decay in a Non--Steady Disk}

For simplicity, above we calculated the timescales $t_{\nu}$ and
$t_{\rm s}$ in steady thin disk models.  However, as noted above, this
highly idealized model makes several crucial assumptions.  In
particular, \citet[hereafter IPP]{ivanov} considered the
tidal--viscous interaction of an unequal mass binary ($q\ll 1$) with a
{\em time-dependent} accretion disk. They assumed that the accretion
disk is initially described by the steady--state solution for a single
BH, and then considered the modifications due to tidal torques from a
secondary BH. The torques are turned on suddenly at some moment $t_0$,
when the secondary, whose mass is $M_2=q M_1 \approx \mu \ll M_1$, is
at an orbital radius $r_0$ that encloses a disk mass $M_{d0}\gsim
M_2$.  The torques are assumed to be concentrated in a narrow ring
near the secondary's orbit, which results in a pile--up of material
near the outer edge of the disk cavity. They found (see their eq. 58)
that this results in a decay time--scale of
\begin{eqnarray}\label{e:t_IPP}
t_{\rm IPP} &=& - \frac{r}{\dot r_{\rm IPP}} = \left(\frac{\mu}{2\dot{M}}\right)
\left(\frac{r_b}{r_0}\right)^{1/2} \tau^{-(a+1)/(2c)}
= \left(\frac{t_{\nu,0}}{q_{B,0}}\right)\left(\frac{r_b}{r_0}\right)^{1/2} \tau^{-(a+1)/(2c)},
\end{eqnarray}
where $\dot{M}$ is the initial steady--state accretion rate,
$t_{\nu,0}$ is the initial viscous time (at $t=t_0$), $q_{B,0}$ is the
disk dominance parameter at $t=t_0$ (see eq.~\ref{e:q_B}), $r_b\equiv
r_b(t)$ is the time-dependent position of the secondary and
$r_0=r_b(t_0)$ is its initial position, and $r_b/r_0 = [1-\gamma
S(\tau^{(5c+b)/(4c)} -1 )]^2$, with the dimensionless time
$\tau=2\beta_0 (t/t_{\nu})$, implying that
\begin{equation}\label{e:tau_IPP}
 \tau = \left\{ \frac{1 + \gamma S - \sqrt{r_b/r_0} }{\gamma S} \right\}^{4c/(5c+b)},
\end{equation}
where $S=(\dot M/\mu) t_{\nu,0}=q_{B,0}/2$, and $a$, $b$, and $k$ are
defined such that $\nu=k \Sigma^a r^b$, $c=2(a+1)-b$,
$\beta_0=[c(2c+a)/[2(a+1)]][(2c+a)/(2c+1)]^{-(2c+a)/(a+1)}$,
$\gamma=2c/[\beta_0(5c+b)]$, and $t_{\nu}$ is the unperturbed viscous
timescale given by equation~(\ref{e:Armitage_Natarayan}) and
calculated explicitly below.\footnote{In IPP, $t_{\nu}$ refers to the
  standard gas pressure dominated accretion disk, which we generalize
  to radiation--dominated disks below.}  If the opacity is dominated
by electron scattering, then $a=2/3$, $b=1$, $c=7/3$, $\beta_0=1.126$,
and $\gamma=0.327$, while for the free-free process $a=3/7$,
$b=15/14$, $c=25/14$, $\beta_0=0.726$, and $\gamma=0.492$. From
equations~(\ref{e:t_IPP}) and (\ref{e:tau_IPP}) we find
\begin{equation}\label{e:t_Ivanov}
t_{\rm IPP} = \left(\frac{t_{\nu,0}}{q_{B,0}}\right)\left(\frac{r_b}{r_0}\right)^{1/2}  
\left\{ \frac{\gamma q_B}{2 + \gamma q_B - 2\sqrt{r_b/r_0} } \right\}^{2(a+1)/(5c+b)}.
\end{equation}
Here, the radial--evolution given by equation~(\ref{e:t_Ivanov}) is
(at least initially) not a simple power--law. Most importantly, as
noted by \citet{ivanov}, the pile--up of the disk material causes the
binary decay to slow--down {\it even more} than estimated for a steady
disk based on the ``disk--dominance'' parameter (eq.~\ref{e:t_II}
above).

\begin{figure}[tbh]
\centering
\mbox{\includegraphics[width=12.5cm]{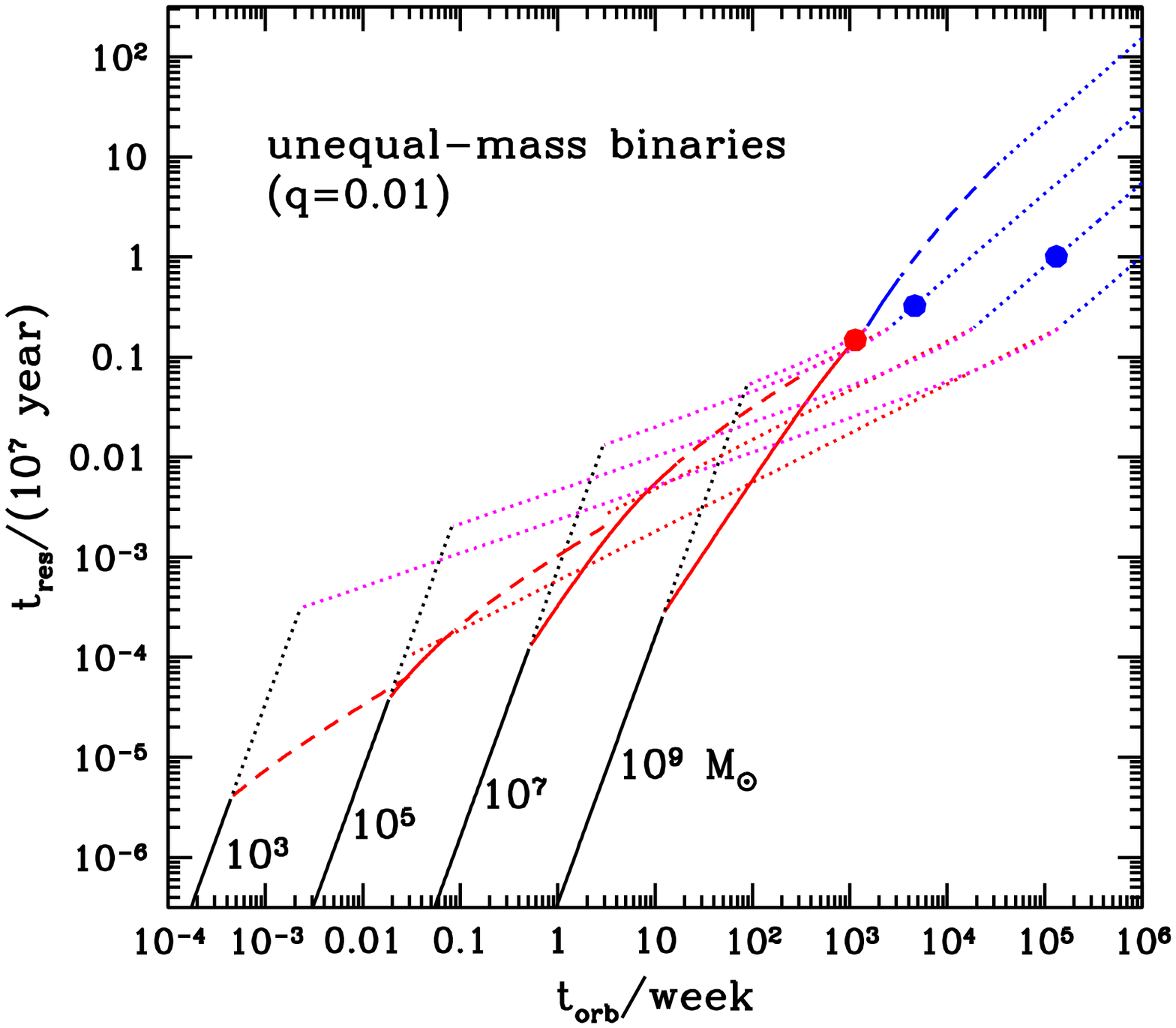}}
\caption{\label{fig:tres_ipp_q1} The residence time for equal--mass
  binaries, as in Figure~\ref{fig:tres_q1}, but with additional curves
  showing the evolution expected in a scenario with a time--dependent
  disk. The dotted magenta curves were calculated based on the model
  by \citet{ivanov}, assuming that the binary--disk interaction turns
  on when the disk dominance parameter reaches $q_B=1$ (prior to this,
  the steady--disk solution is applied).  As expected, the pile--up of
  material in the time--dependent disk slows down the decay of the
  binary.}
\end{figure}

\begin{figure}[tbh]
\centering
\mbox{\includegraphics[width=12.5cm]{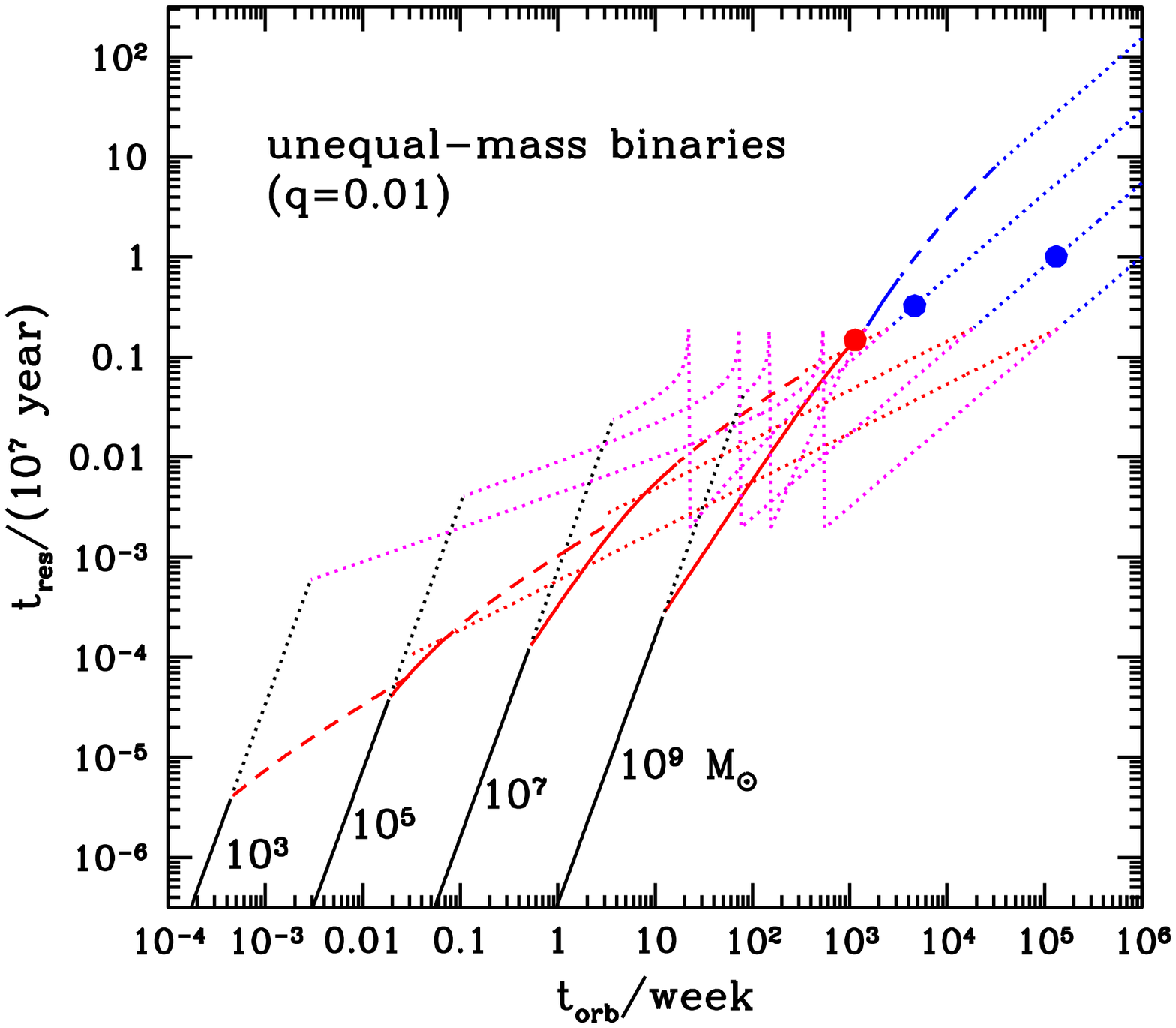}}
\caption{\label{fig:tres_ipp_q0.01} The residence time $t_{\rm res}$
  as in Figure~\ref{fig:tres_ipp_q1}, except that the binary--disk
  interaction, as modeled by \citet{ivanov}, is turned on later, when
  the disk dominance parameter drops below $q_B=0.01$. Prior to this,
  the disk--dominated, steady--disk solution is applied.}
\end{figure}

In Figures~\ref{fig:tres_ipp_q1} and \ref{fig:tres_ipp_q0.01}, we
illustrate the impact of allowing the disk to evolve.  In the above
approach of \citet{ivanov}, we have to specify when the interaction
between the secondary and the disk is turned on.  In
Figure~\ref{fig:tres_ipp_q1}, we assume that the interaction begins as
soon as the disk dominance parameter reaches $q_B=1$. In
Figure~\ref{fig:tres_ipp_q0.01}, we delay the onset of the interaction
to $q_B=0.01$.  In both figures, the new dotted (magenta) curves
denote the binary's residence time in the time--dependent disk.  Note
that in the latter case, in Figure~\ref{fig:tres_ipp_q0.01}, the
residence time undergoes a discrete jump when the disk--binary
interaction is turned on: the binary stalls, and does not move
initially, until the mass of material that has piled up is of the
order of the secondary's mass.  As the figures show, these residence
times are indeed significantly longer than in the steady--disks.  At
relatively late times after the interaction is assumed to turn on, the
residence times asymptote to power--law forms. Most significantly, for
each of the binaries shown in Figures~\ref{fig:tres_ipp_q1} and
\ref{fig:tres_ipp_q0.01}, the transition to the GW--driven regime
occurs significantly earlier due to the evolution of the disk; just
before this transition to the GW--driven regime, the residence times
are longer by $\approx$ two orders of magnitude compared to a steady
disk.

\section{Observational Search for SMBH Binaries Among Luminous AGN}
\label{sec:observations}

In the rest of this {paper}, we will discuss identifying coalescing
SMBHBs with quasars, and interpreting the residence time $t_{\rm res}$
as the duty--cycle for exhibiting periodic variability on the observed
time--scale $t_{\rm var}\approx (1+z)t_{\rm orb}$.  Our broad
justification for these hypotheses is the generic idea, advanced in
numerous other works, that quasars are activated in major galaxy
mergers (e.g. \citealt{hopkins07a} and references therein). Since
SMBHs are believed to be common in galactic nuclei (at least at low
redshifts $z\lsim 3$; see \citealt{mhn01,lippaimergers} and references
therein) there could then arguably be a one--to--one correspondence
between the quasar phenomenon and SMBHB coalescences.

\subsection{A Simplified Model for the Population of Periodic AGN}
\label{sec:periodic}

The total quasar lifetime, defined as the cumulative duration
(possibly over multiple episodes) for an individual source to produce
bright emission near the Eddington limit, is generally believed to be
$t_{\rm Q}\approx {\rm few} \times 10^7$ yr, based on several lines of
observational evidence \citep{martini04}.  As discussed above, this
value is consistent with the time--scale it takes for a binary SMBH to
evolve to coalescence, starting from the outer edge of a
gravitationally stable thin $\alpha$--disk.  Therefore, we hypothesize
that the luminous quasar phase coincides with this last stage in the
merger of the two SMBHs.  Of course, it is possible that the quasar
phase occurs either {\it long before} or {\it after} coalescence -- in
either case, there would be no bright emission to observe during the
last stages, as hypothesized here.\footnote{For example, the model by
  \citet{co01} for episodic quasar activity involves a single SMBH,
  and does not require the presence of a binary.}

We next assume that during the coalescence, the binary produces a
steady luminosity $\bar{L}_{\rm Q}$ (which evolves only on long
time--scales $\gg t_{\rm orb}$), with roughly periodic fluctuations of
amplitude $\Delta L_{\rm Q}$ and period $t_{\rm var}=(1+z)t_{\rm orb}$
about this steady mean luminosity.  As argued in the Introduction,
periodic variations could be reasonably expected if the luminosity is
tied to the mass accretion rate, with the latter modulated on the
orbital period. Even in the absence of such modulations, the emission
could vary owing to the orbital motion and emission geometry of the
binary (Kocsis \& Loeb, in preparation).  In the absence of a
quantitative model for the electromagnetic emission, we will assume
the amplitude $\Delta L_{\rm Q}$ is unknown, and below we will ask
whether a particular assumed $\Delta L_{\rm Q}$ may be detectable.
Since the residence time $t_{\rm res}$ decreases continuously as the
binary separations shrinks, variability with decreasing periods
$t_{\rm orb}$ would be exhibited by a diminishing fraction $\sim
t_{\rm res}/t_{\rm Q}$ of bright quasars.

{\em Our main point is that an observational survey can attempt to
identify such periodically variable sources.} The total number of such
periodic sources will be $N_{\rm var}\approx t_{\rm res} \dot{N}_{\rm
mg}$, where $\dot{N}_{\rm mg}$ represents the merger rate between BHs
within the survey volume (or more precisely, the activation rate of
SMBH coalescence events).  In general, the merger rate depends on
redshift and on both BH masses, or $\dot{N}_{\rm mg}=\dot{N}_{\rm
mg}(z,M,q)$, and should include only those sources with a luminosity
above the survey detection threshold.  To account for the latter
condition, the light--curve of each SMBHB, $L_{\rm Q}=L_{\rm
Q}(t,M,q)$ needs to be known (here $t$ could, for example, refer to
the look--back time before merger).

The merger rate $\dot{N}_{\rm mg}(z,M,q)$ can be modeled using the
dark matter halo merger rate with a recipe of associating BHs with
halos, and the light--curve $L_{\rm Q}(t,M,q)$ can then be constrained
by matching the observed quasar luminosity function
\citep[e.g.][]{kh00}.  However, a large range of such BH population
models can fit the observational data
\citep[e.g.][]{mhn01,lippaimergers}.  To proceed, we instead make the
simple assumption that each BH binary produces a constant mean
luminosity of $\bar{L}_{\rm Q}=f_{\rm Edd}L_{\rm Edd}$ for a total
duration $t_{\rm Q}$ during its lifetime, where $f_{\rm Edd}$ is a
constant of order unity.  Reasonable fiducial values appropriate to
the bright quasar phase are $f_{\rm Edd}\approx 0.3$
\citep{kollmeier06} and, as mentioned above, $t_{\rm Q}\approx {\rm
  few} \times 10^7$ yr \citep{martini04}.  Note, in particular, that
the quasar lifetime $t_{\rm Q}$ is known to be much shorter than the
Hubble time, and $\dot{N}_{\rm mg}$, which is likely determined by the
galaxy merger rate, and proceeds on a cosmological time scale, can
reasonably be assumed to be constant during $t_{\rm Q}$.

Under the above assumptions, the fraction $f_{\rm var}$ of objects
with luminosity $L_{\rm Q}$ that display periodic variability on the
time-scale $t_{\rm var}$ is simply given by the ratio $f_{\rm var} =
t_{\rm res}/t_{\rm Q}$.  This ratio can be read off directly from
Figures~\ref{fig:tres_q1}-\ref{fig:tres_ipp_q0.01}. Note that this
conclusion still holds if the quasar emission is intermittent; we
require only that the quasar is ``on'' for the duration $t_{\rm res}$
when the binary orbital timescale is $t_{\rm orb}$.  Most importantly,
under these assumptions, the predicted number $N_{\rm var}= (t_{\rm
  res}/t_{\rm Q}) N_{\rm tot}$ is a fixed fraction of the total number
$N_{\rm tot}$ of quasars, and is independent of the merger rate, as
long as the latter is constant during $t_{\rm Q}$. We can then
associate $N_{\rm tot}$ with the {\it observed} number of bright AGN.
In particular, in the GW--dominated regime, we have the simple
prediction
\begin{equation}
\label{eq:Nvar}
f_{\rm var} = \frac{N_{\rm orb}}{N_{\rm tot}} =
\left(\frac{10^7 {\rm yr}}{t_Q} \right)
\left[\frac{t_{\rm var}}{50.2 (1+z)\, {\rm week}} \right]^{8/3}
M_{7}^{-5/3}
q_s^{-1}.
\end{equation}
Note that in this equation, $t_{\rm var}=(1+z)t_{\rm obs}$ is the
variability time-scale as observed on Earth (assumed to equal the
redshifted orbital time); the quasar lifetime $t_Q$ is evaluated in
the quasar's rest--frame.

Before we proceed, we emphasize that there are many complications over
the above, simplified picture.
First, luminous quasar activity requires a near--Eddington
mass--accretion rate, with the gas reaching within several
Schwarzschild radii of one or both BHs.  It is unclear whether
abundant gas will indeed be present this close to the BHs, especially
since during the late stages of the merger, the gas is evacuated from
the inner disk by the binary's torques, and the exterior gas disk is
eventually unable to follow the rapidly decaying BH binary.
Furthermore, in the final, GW--dominated regime, the $t_{\rm
  res}\propto t_{\rm orb}^{8/3}$ scaling strictly holds only if any
residual circumbinary gas has negligible impact on the orbital decay.
This requirement could, in fact, contradict the assumption that the
binary is producing bright emission during this stage.  Second, in
order for the emission to be periodically variable, the gas has to
respond rapidly to the gravitational perturbations from the
binary. The time--scale for this response is of order the {\it local}
orbital time; variability on the orbital time--scale of the {\it
  binary} itself therefore again requires gas close to the binary's
orbital radius.

If the central cavity were indeed truly empty, no gas would reach the
SMBHBs, and bright emission could not be produced.  On the other hand,
an empty cavity is certainly an idealization, and detailed models for
the joint disk + binary evolution are required to assess the
plausibility of our assumptions. Conversely, the observations
envisioned here will constrain such models (which, again, is the
main point of the present paper).

In support of our assumptions, we note, however, that gas could be
present near the BHs in the case of unequal masses (so that the
torques are reduced), or if the disk remains thick, making it
difficult for the binary to open and maintain a nearly empty central
cavity.  Numerical simulations indeed suggest residual gas inflow into
the cavity \citep{al96,mm08,hayasaki07,cuadra09}, which may plausibly
accrete onto the BHs (with both BHs possibly forming their smaller
individual accretion disks; \citealt{hayasaki08}), producing
non--negligible EM emission. Simulations have also shown, in the
context of proto--planetary disks, that when the circumbinary disk is
sufficiently thick, the mass flow rate across the gap is increased
\citep{dobbs07}. Such residual inflow onto a SMBH binary has been
invoked to explain the $\sim$12--yr periodic emission from the quasar
OJ287 \citep{al96}. More recently, the large velocity offsets seen in
the spectrum of the quasar SDSS~J092712.65+294344.0 \citep{komossa08}
have been interpreted with a similar model, including gas inflow onto
a luminous SMBH binary \citep{bogdanovic09}; a similar interpretation
was invoked for the binary quasar candidate recently identified by
\citet{boroson09}.

There are additional caveats that will hamper the identification of
any periodic sources, even if they exist and produce bright enough
luminosity to be detectable.  The Eddington ratio of bright AGN is
already known to have a significant scatter ($\sim 0.3$ dex;
\citealt{kollmeier06}).  The light--curve of the merging binary is
also likely to evolve, rather than having a simple ``tophat'' shape.
It is possible, in particular \citep[e.g.][]{barger05,hopkins05} that
merging SMBHs spend a significantly longer time ($\sim 10^9$ yr) at
lower luminosities, ($f_{\rm Edd}\ll 1$).  This will complicate the
interpretation of any observed variability (i.e., converting the
observed ratio $N_{\rm var}/N_{\rm tot}$ at $t_{\rm var}$ to $t_{\rm
res}$ will require knowing the probability distribution of Eddington
ratios).  This, however, can be alleviated by considering only the
{\it relative} abundance of periodically variable objects at different
values of $t_{\rm var}$, instead of the absolute number of sources
that show periodic variability. In this case, the only assumption
required is that $f_{\rm Edd}$ does not evolve significantly during
the observed range of $t_{\rm var}$ -- this should be reasonable over
a factor of a $\sim$ few range in orbital radius or in $t_{\rm orb}$.
Furthermore, even if there is a range of different BH masses, among
sources with a similar luminosity, producing variability with the same
period, Figures~\ref{fig:tres_q1} and \ref{fig:tres_q0.01} show that
more massive BHBs will move much more quickly through a fixed $t_{\rm
orb}$. Given that there are most likely fewer of the more massive BHBs
to begin with, the set of all sources with the same $t_{\rm var}$ will
be heavily dominated by the lowest--mass BHBs, caught at their
relevant orbital radius.  This still leaves the caveat, however, that
the source is significantly sub--Eddington during the late stages of
coalescence.  In this case, the periodic sources will be harder to
detect both because they are fainter, and also because they will also
be rarer (among the long--lived and therefore more numerous,
near--Eddington quasars with a similar luminosity).

Another caveat is that at fixed $t_{\rm orb}$ and $M$, the
distribution of $q$ is unknown, and can depend on $M$.  However,
bright AGN activity is thought to be activated only in relatively
major mergers. A smaller satellite galaxy, falling onto a larger
central galaxy that is more than $\approx 10$ times more massive, may
not experience the torques needed to bring its gaseous nucleus, with
the low--mass BH, close to the center of the larger galaxy, for the
BH-BH merger to take place \citep{hopkins06}; the dynamical friction
time for small galaxies themselves can also be too long
\citep[e.g.][]{kh00}, and/or the small satellites can be tidally
stripped before reaching the central regions of the larger galaxy
\citep[e.g.][]{kazantzidis05}. These arguments, coupled with the
well--established correlations between the mass of a SMBH and its host
galaxy (e.g. \citealt{ferrarese02}, see also the Introduction),
suggest that the $q$--distribution among binaries associated with
quasars may not extend to values significantly below $q\sim 0.1$.

Finally, for simplicity, in our estimates we have assumed circular
orbits, both for the binary and the disk gas.  It has been shown that
the binary--disk interaction could drive both the SMBHs and the gas to
have significant eccentricities
\citep{an05,mm08,dot08,hayasaki08,cuadra09}.  Such eccentricities
should leave characteristic asymmetric signatures in the modulated
mass accretion rate (see Figure 8 in \citealt{hayasaki07}). The
resulting light--curves may exhibit corresponding features, which
could be resolved, given sufficient time--sampling. In practice,
allowing for eccentricities will most likely further complicate the
interpretation of any observed period distribution, especially if the
time--sampling is too coarse to explicitly reveal any asymmetric
features.

\subsection{Requirements of a Variability Search}

Despite the caveats listed in the previous section, it is plausible
that the periodic sources envisioned here exist, and we propose that
they can be looked for, in a suitably designed survey.  Most
importantly, Figures~\ref{fig:tres_q1} and \ref{fig:tres_q0.01} show
that the expected variability timescale can be in a suitable range for
a statistical detection, with a duty--cycle of $t_{\rm res}\gsim 10^4$
yr over the range from $t_{\rm var}\sim$day to $\sim$yr.  This
suggests that such periodic sources may not be too rare.

What will be the practical limitations for discovering the population
of periodic sources?  Clearly, there has to be a sufficient number of
sources, observed over a range of variability time--scales for a
representative statistical sampling, and the brightness variations of
these sources must be detectable. In addition, the individual
light--curves have to be sampled well enough to confirm their periodic
nature: this will be necessary to distinguish the coalescing SMBH
binaries from other types of variable objects.  Besides discovering
the periodic sources, the idea proposed here is to measure the
dependence of $N_{\rm var}$ on $t_{\rm var}$ -- possibly to use the
$N_{\rm var}\propto t_{\rm var}^{8/3}$ scaling to demonstrate that the
periodic variability comes from perturbations by the orbital motion
during the GW inspiral. For this, the survey also needs to cover at
least a factor of several range in $t_{\rm var}$.

The above issues will place requirements on (i) the sensitivity and
(ii) solid angle, as well as on the (iii) total duration and (iv)
sampling rate for a survey.  We can use the simple disk models and the
idealized picture discussed above, to roughly delineate these
requirements.  For simplicity of discussion, let us assume that all
sources are at $z=2$.  In reality, quasars (and therefore major BH
mergers) have a broad distribution with a peak around this redshift;
clearly this will have to be taken into account in designing an actual
survey.  For simplicity, let us also fix the mass ratio $q=1$. In
reality, there should be a distribution of values, perhaps in the
range $0.1\lsim q \leq 1$, for the mergers that activate bright quasar
activity \citep{hopkins06}. This would not significantly affect our
conclusions, unless $q$ frequently extends well below 0.1.

Imagine a survey with a sensitivity that corresponds to detecting the
periodic variability of BHBs with a mass $M_{\rm min}$ at $z=2$,
covering a solid angle $\Delta\Omega$. (A real survey, of course, will
have a completeness for variability detection that is not a step
function).  Let us assume that the variable flux corresponds to a
fraction $\eta_{\rm var}$ of the steady mean luminosity, $\Delta L_{\rm
Q}=\eta_{\rm var}\bar{L}_{\rm Q}=\eta_{\rm var}f_{\rm Edd}L_{\rm Edd}$.  If
the survey volume contains a total of $N_{\rm tot}$ SMBHBs with the
luminosity $\bar{L}_{\rm Q}$, then the periodic variable fraction,
$t_{\rm res}/t_{\rm Q}$, can be determined down to the smallest value
$\approx N_{\rm tot}^{-1}$ (i.e. to find at least one periodic
source).  Fixing the values of $\eta_{\rm var}f_{\rm Edd}$ and $t_{\rm
Q}$ (as well as $M_{\rm min}$, $z$ and $q$), this corresponds to a
minimum variability time--scale $t_{\rm var, min}$ that can be probed.
Let us define the requirement that this minimum is $t_{\rm var,
min}\leq 20$ weeks.  Assuming that the longest variability time--scale
of interest is around $t_{\rm var, max} \sim 1$ year (so that the
periodic nature of the variations can be convincingly demonstrated
over a multi--year survey), this will offer a factor of three range in
$t_{\rm var}$ for mapping out the $N_{\rm var}$ {\it vs.} $t_{\rm
var}$ dependence. For example, with the steepest possible (pure
GW--driven) scaling $N_{\rm var}\propto t_{\rm var}^{8/3}$, a survey
volume containing a single source with $t_{\rm var}=20$ weeks would
then contain $3^{8/3}\approx 20$ sources with a similar luminosity but
with a $t_{\rm var}=60$ week period.

\begin{figure}[tbh]
\centering
\mbox{\includegraphics[width=12.5cm]{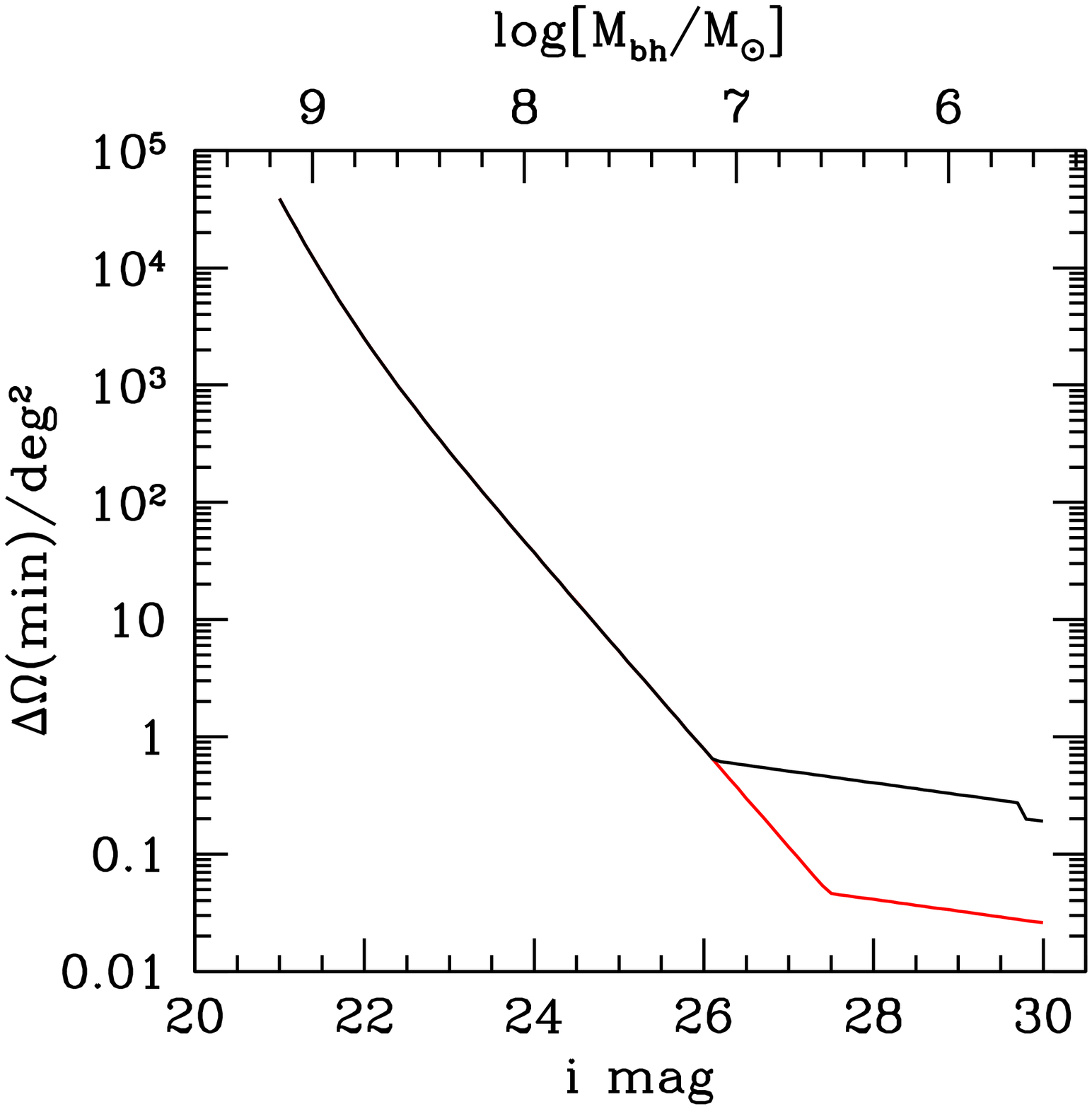}}
\caption{\label{fig:survey} The sky coverage required to find a
  population of periodic sources, assuming our fiducial set of source
  parameters (see text).  The upper (black) curve shows the solid
  angle required to find at least $\approx 100$ periodic sources
  between $1.5<z<2.5$ with an observed period of 60 weeks, as a
  function of the limit of the survey for the variable $i$--band
  luminosity.  The lower (red) curve shows the solid angle required
  for at least 5 periodic sources with $t_{\rm var}=20$ weeks.  In the
  GW--driven regime, these two criteria coincide.  The upper labels
  show the mass of the SMBHB producing the corresponding steady $i$
  magnitude (assumed here to be 2.5 mag brighter than the variable
  magnitude).  As shown by the break between 26 - 27 mag in the
  curves, BHs with a mass above/below $\sim 10^7 {\rm M_\odot}$ are in
  the GW/gas--driven regime, respectively.}
\end{figure}

To fix some numbers, let us set $f_{\rm Edd}=0.3$, $\eta_{\rm var}=0.1$,
and $t_{\rm Q}=10^7$ yr.  For reference, the Eddington luminosity of a
$3\times 10^6 \,{\rm M_\odot}$ BH at $z=2$, assuming a $\sim$ 10\%
bolometric correction, corresponds to an optical magnitude of
$\approx$24 mag (in the $i$ band).  Let us also impose the (somewhat
ad--hoc) requirement that the survey volume should contain at least
$N_{\rm var}\geq 100$ sources with a detectable flux variations at the
period of $60$ weeks.  In the GW--driven stage, there will then be at
least 5 detectable periodic sources with a period of $\leq 20$ weeks;
in the gas--driven regime, where the scaling $f_{\rm var}$ {\it vs.}
$t_{\rm var}$ is flatter, there will be a larger number of $20$--week
period sources.

In Figure~\ref{fig:survey}, the curves show the sky coverage required
to satisfy these criteria, as a function of the $i$--band variable
magnitude corresponding to the detection limit of the survey.  The BH
masses producing the corresponding steady $i$ magnitude (which, in our
fiducial model, is 2.5mag brighter than the variable magnitude) are
shown on the top axis.  This figure assumes $q=1$.  We used the
fitting formula by \citet{hopkins07b} for the bolometric quasar
luminosity function (LF) $d\phi/dL(z,L)$ to compute the the total
number $N_{\rm tot}$ of quasars at $z=2$, per solid angle
$\Delta\Omega$, in a redshift range of $\Delta z=1$, i.e. $N_{\rm
  tot}=(\Delta z \Delta\Omega) (d^2V/dzd\Omega) \int_{L_{\rm
    min}}^\infty (d\phi/dL)dL$, where $(d^2V/dzd\Omega)$ is the
cosmological volume element, and $L_{\rm min}$ is the bolometric
luminosity corresponding to the steady magnitude threshold $i$.  We
then used equation~(\ref{eq:Nvar}) for $f_{\rm var}$ to obtain the
total number $N_{\rm var}=f_{\rm var}N_{\rm tot}$ of variable sources
at observed period of $t_{\rm var}=20$ weeks and at $t_{\rm var}=60$
weeks.  Requiring $N_{\rm var}$($t_{\rm var}=60$ weeks)$\geq 100$ then
yields the solid angle $\Delta\Omega$ as a function of $i$.  Note that
the quasar LF is almost a pure power--law up to $M_{\rm bh}\approx
10^9~{\rm M_\odot}$.  The break between $i$ = 26 - 27 mag in the solid
curves corresponds to the transition between GW and gas--driven
orbital decay.  In particular, the figure shows that SMBHs with a mass
above/below $\sim 10^7 {\rm M_\odot}$ are in the GW/gas--driven
regime, respectively.

Figure~\ref{fig:survey} shows that there is a clear trade--off between
survey depth and area: the required sky coverage scales with the
survey flux limit approximately as $\Delta\Omega \propto F^{-2}$, with
a steepening for shallow surveys with limiting magnitudes $i\lsim
22.5$ (due to the decline at the bright end of the quasar LF), and a
flattening for very deep surveys with limiting magnitudes $i\gsim
26.5$ (because the SMBHBs are in the gas--driven regime and their
residence times at fixed $t_{\rm orb}$ are shorter than in the pure
GW--driven regime).

From Figure~\ref{fig:survey}, we conclude that, for example, a
1~sq. degree survey, detecting SMBHBs whose steady luminosity is
$i=23.3$ mag, with a variability at the level of $i=25.8$ mag, with
sufficient sampling and duration to cover periods of 20--60 weeks,
represents an example for the minimum specification for the survey
parameters (i)-(iv).  In this example, the mass of the BHs being
detected is $\sim 2\times 10^7~{\rm M_\odot}$.
Figures~\ref{fig:tres_q1} and \ref{fig:tres_q0.01} show that at
orbital periods of $(20-60)/(1+z)=7-20 $ weeks, these BHs are all in
the GW--driven regime when $q\sim 1$, but may be in the
viscosity--driven regime for $q\ll 1$.  Surveys that go deeper and
cover a smaller area will begin probing the gas--driven evolutionary
stages.

\begin{figure}[tbh]
\centering
\mbox{\includegraphics[width=12.5cm]{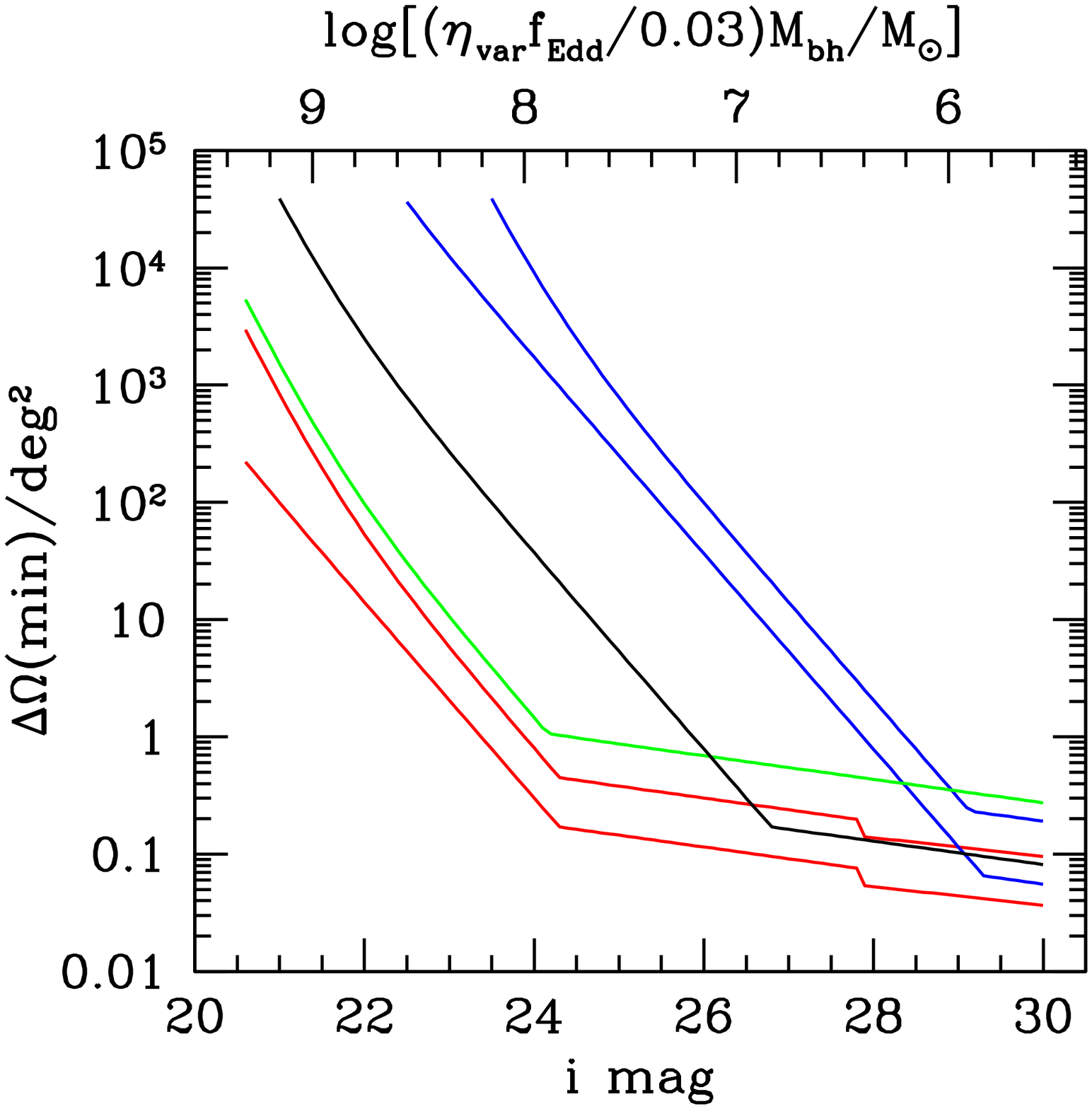}}
\caption{\label{fig:survey2} This figure shows how the sky coverage
  required to find a population of periodic sources depends on
  variations in our fiducial set of source parameters.  The middle
  (black) curve shows the sky coverage required to find 20 sources at
  $t_{\rm var}=35$ weeks, with the same set of fiducial parameters,
  $q=1$, $\eta_{\rm var}=0.1$, $f_{\rm Edd}=0.3$, as in
  Figure~\ref{fig:survey}.  The green curve corresponds to changing
  the mass ratio to $q=0.01$; the top/bottom (blue/red) pair of 
  curves show variations when either $\eta_{\rm var}$ or $f_{\rm Edd}$ is
  decreased/increased by a factor of 10. Note that the survey volume
  requirement is more sensitive to $\eta_{\rm var}$ (whereas the mass of
  the smallest detectable periodically variable SMBHB, shown on the
  top axis, is equally sensitive to either).}
\end{figure}

In Figure~\ref{fig:survey2}, we examine how the required sky coverage
changes when the parameters $q$, $\eta_{\rm var}$ or $f_{\rm Edd}$ are
modified.  The middle (black) curve shows the sky coverage
required to find 20 sources at $t_{\rm var}=35$ weeks (intermediate
between the red and black curves in Figure~\ref{fig:survey}),
with our fiducial parameters, $q=1$, $\eta_{\rm var}=0.1$, $f_{\rm
  Edd}=0.3$.  The green curve corresponds to changing the mass ratio
to $q=0.01$; this increases/decreases the residence time in the
GW/gas--driven regimes relative to the $q=1$ case (compare
Figs.~\ref{fig:tres_q1} and \ref{fig:tres_q0.01}), and therefore
reduces/increases the required solid angle coverage.  The top pair of
(blue) curves in Figure~\ref{fig:survey2} show variations when
either $\eta_{\rm var}$ or $f_{\rm Edd}$ is decreased by a factor of 10
(to 0.03 or 0.01, upper and lower of the pair, respectively). Note
that the survey volume requirement is more sensitive to $\eta_{\rm var}$
(whereas the critical BH mass is equally sensitive to either).
Similarly, the bottom pair of (red) curves show variations when
either $\eta_{\rm var}$ or $f_{\rm Edd}$ is increased by a factor of 10.
The small break visible at $\approx25.5$ mag in this case corresponds
to the transition from the middle to the outer disk region for
$\approx 2\times 10^5~{\rm M_\odot}$ SMBHBs at $t_{\rm
  orb}=35/(1+z)\approx 11.5$ weeks (see Fig.~\ref{fig:tres_q1}).  Each
dashed curve shows the mass of the SMBHB corresponding to the
effective $i$ magnitude limit (BH masses are labeled on the right
$y$-axis).  The required survey volume also shifts linearly with the
assumed total quasar lifetime $t_{\rm Q}$.

If we knew where to look (i.e., if {\it LISA} delivers a candidate for
an on--going merger, with sufficiently accurate localization on the
sky), it would be possible to perform a deep, targeted observation for
variability on short time--scales; between several minutes up to $\sim
10$ hours within the last $\sim$ month of merger \citep[this
possibility is discussed in detail in][]{paper2}. However, each source
will spend only a $\sim$month at such short variability time--scales,
and a random search, in the absence of a preferred direction on the
sky, would then have to monitor $>10^8(t_{\rm Q}/10^7{\rm yr})$ AGN to
find a single example of such a late--stage periodic source.
Alternatively, one may monitor $\sim 10^6(t_{\rm Q}/10^7{\rm yr})$ AGN
for $\sim 10$ years, to look for (slowly evolving) periods, on the
timescale of $\sim$ a day.  The slow decrease in the period, which
would be a smoking gun for GW--inspiral, will be challenging to
observe in real time for individual objects.

\subsection{Constraints from Existing and Future Data}
\label{sec:existing}

Existing observations from radio to X--ray bands have shown that the
luminosity of quasars and other active galactic nuclei varies on
time--scales from hours to several years (see, e.g., the articles in
\citealt{variabilitybook} or the recent review by
\citealt{fanjh05}). In fact, variability often aids in the
identification of AGN (and may conversely be a major obstacle in
identifying the periodic signal proposed here).  While variability is
detected in a large fraction of all AGN, there are only a handful of
sources whose structure function shows clear {\em periodic}
variability on long ($\gsim$ weeks) time--scales (see,
e.g. \citealt{rieger07} for a review focusing on searches for periodic
variability). Examples include a handful of blazars, whose historical
light--curves show periodic outbursts on timescales of a year to a
decade, or even longer \citep[see, e.g.,][and references in these
papers, for individual
objects]{sillanpaa,liu95,liu97,raiteri01,qian07,tao08}.
\citet{neugebauer} monitored 25 low--redshift, optically selected
quasars for variability over several decades in infrared bands. They
identified one object, the radio--loud quasar PG 1535+547, whose
structure function shows a periodic component with a period of $\sim
10$ yr.  This source has a bolometric luminosity of $L_{\rm
  bol}=10^{13.44}{\rm L_\odot}$, implying a BH mass of $\sim 3\times
10^9~(f_{\rm Edd}/0.3)^{-1} {\rm M_\odot}$. Figure~\ref{fig:tres_q1}
shows that equal--mass SMBH binaries with this mass are in the
GW--driven regime, and $f_{\rm var}\sim 20\%$ may exhibit a 10--year
period. In comparison, Figure~\ref{fig:tres_q0.01} shows that
unequal--mass binaries with this total mass may be in the gas--driven
regime, and periodic variability may be exceedingly rare.  Thus, we
conclude that the identification of one periodic object is roughly
consistent with it being an example of a GW--driven, near--equal mass
binary. However, there are only 4 objects in the sample studied by
\citet{neugebauer} with luminosities above $L_{\rm bol}=10^{13}{\rm
  L_\odot}$, prohibiting robust conclusions.

Very large area variability surveys, such as in the Sloan Digital Sky
Survey (SDSS), are shallow, and detect variability down to only $i
\approx 20$ mag~\citep{vandenberk04}.  With our fiducial $f_{\rm
var}=0.1$, this corresponds to periodic AGN whose steady luminosities
are $i \approx 17.5$ mag. Figure~\ref{fig:survey} shows that such a
survey would have missed the periodic variations discussed above.
Figure~\ref{fig:survey2} shows that if the variable fraction is much
larger, $\eta_{\rm var}\gsim 0.3$, then $\sim 100$ variable sources with
a yearly period would be detectable -- but their periodic nature could
be demonstrated only with sufficient time--sampling, extending over a
decade.

A recent, much deeper optical survey by the Subaru telescope
\citep{morokuma1,morokuma2} for variable objects provides interesting
constraints on the scenario envisioned here.  The completeness
function in this survey, defined as the probability to detect flux
variations of an object with a variable component $i$, goes from
$\sim$unity to $\sim$zero between $i\approx 25$ to $i\approx 26$ (see
Fig. 8 in \citealt{morokuma1}). In the fiducial case with $f_{\rm
  Edd}=0.3$ and $\eta_{\rm var}=0.1$, the limiting variability magnitude
$i=25.5$ mag corresponds to the mean steady magnitude of $i=23$ mag,
and BH mass of $M=2.5\times 10^7~{\rm M_\odot}$.  At $z=2$, the Subaru
survey has a completeness of 0.5 at this magnitude (see Fig. 11 in
\citealt{morokuma1}) and covers an area of 0.9 sq. degrees.  This
combination of sensitivity and area lies very close (just below) the
curves in Figure~\ref{fig:survey}. Using the \citet{hopkins07b} quasar
LF, and assuming a completeness of 0.5, we find that the Subaru survey
should detect $440$ AGN; this is in nearly exact agreement with their
quoted result (489 deg$^{-2}$).  We further find that of these
sources, $\sim 0.6, 12, 61$ would vary with observed periods of 20
weeks, 60 weeks, and 1000 days (adopting a probability of 0.5 for
detecting variability, from Fig. 8 in \citealt{morokuma1}).  Figure 12
in \citet{morokuma2} shows that they found several dozen sources that
varied, at least once in their life, on all of these timescales.
Unfortunately, we do not know whether these sources are periodic or
not, and therefore the Subaru survey results represent only an upper
limit on the fraction of periodic sources. Nevertheless, this already
suggests that the $2.5\times 10^7 {\rm M_\odot}$ BHs at the limit of
the survey can not produce variability at the level significantly
exceeding our fiducial $0.03 L_{\rm Edd}$.

The ultra--deep {\it Hubble Space Telescope (HST)} variability surveys
\citep[see, e.g., the recent review by][and references
therein]{sarajedini} discovered galaxies whose nuclei varied by
magnitudes down to $V\approx 27.5$. The observations were taken a year
apart in the Hubble Deep Field North (HDFN) and the Groth Survey Strip
(GSS), whose areas are $\sim 3\times 10^{-3}$ and $\sim 0.1$
sq. degrees, respectively.  While the solid angle of the HDFN dataset
is too small to yield useful constraints, the fiducial case with
$f_{\rm Edd}=0.3$ and $\eta_{\rm var}=0.1$ in Figure~\ref{fig:survey}
shows that the GSS dataset just reaches the sensitivity/area
combination of $\sim 27$ mag and $\sim 0.1$ sq. degrees required to
find flux variations from $M_{\rm bh} \sim 10^7 {\rm M_\odot}$ SMBHBs.
Approximately $4.5\%$ of AGN were found to vary by magnitudes down to
$V\approx 27$ in this dataset \citep[see][for more
details]{sarajedini06}, suggesting that $\lsim 5\%$ of AGN containing
SMBHBs with this mass can produce variability at the $\sim 0.03 L_{\rm
Edd}$ level.

AGN are also known to vary on long times--scales in X-ray bands.
Systematic and unbiased variability surveys sensitive to times--scales
of weeks, such as those in soft X--rays in the ROSAT all sky survey
\citep{fuhrmeister} or in hard X--rays in {\it Swift}/BAT data
\citep{beckmann} however, have been restricted to the brightest AGN,
while deeper surveys, such as those by {\it RXTE} \citep{markowitz},
of the {\it Chandra} Deep Field North \citep{bauer03,bauer04} and
South \citep{paolillo04}, and by {\it XMM} \citep{papadakis08} have
only monitored up to a few hundred sources.  These observations do
suggest that a large fraction of AGNs vary in X--ray bands on time
scales of a day to a year, but whether the variations are periodic
have not been determined.  In the 9--month duration observations
covered by {\it Swift}/BAT data, \citet{beckmann} find a strong
anti--correlation between luminosity and variability (with no source
with luminosity $L_X>5\times 10^{43}~{\rm erg~s^{-1}}$ showing
significant variability); \citet{papadakis08} report a similar trend
from an X-ray variability analysis of 66 AGN in the Lockman Hole.
These findings would be consistent with the trend that the most
massive SMBHBs ($\gsim 10^8~{\rm M_\odot}$) spend less time at a fixed
orbital timescale of $t_{\rm orb} \sim 10$ weeks (see
Figures~\ref{fig:tres_q1} and \ref{fig:tres_q0.01}).  The results of
\citet{beckmann} suggest that absorbed sources vary more than
unabsorbed ones, which may be particularly relevant for finding the
periodic SMBH binary sources envisioned here, which are undergoing the
last stages of their merger, and may be heavily obscured and visible
primarily in X--ray bands.

While the deep existing optical surveys come close to placing useful
constraints on the scenario envisioned here, future surveys, designed
to uncover source populations with periodic variations on
times--scales of tens of weeks, should be able to either discover
these populations, or place stringent limits on their existence.  Many
large optical/IR surveys are being planned or built, motivated largely
by finding type Ia supernovae (SNe) for cosmological studies
\citep[see, e.g.,][for a recent review]{stubbs}. The most ambitious of
these, such as LSST and Pan-STARRS-4\footnote{see www.lsst.org and
www.pan-starrs.ifa.hawaii.edu} will be all--sky surveys, and should be
able detect variability to $26-27$mag, allowing detections well beyond
the most pessimistic case shown in Figure~\ref{fig:survey2}.  The
proposed ALPACA survey \citep{alpaca} will cover 1,000 sq.  degrees to
23-25 mag in 5 optical bands, and would already reach the
sensitivity/area combination probing these pessimistic scenarios.

\subsection{Other Detection Methods}
\label{sec:alternatives}

In addition to producing periodic variability, there could be several
other methods to prove or disprove the presence of a SMBH binary.
First, the orbital motion of the binary may cause relative shifts in
the quasar's emission lines.  For example, in a configuration in which
the broad lines arise from gas close to one of the two (moving) BHs,
and the narrow lines arise from material farther away, which is close
to rest at the systemic redshift, such a shift could arise between the
narrow and broad emission lines \citep{bogdanovic09}.  Similarly, if
both BHs carry their own accretion disks, extending to a few
Schwarzschild radii, and produce broad lines \citep{hayasaki07}, then
there could be two sets of broad emission lines, super--imposed with a
similar relative velocity shift.  The magnitude of these shifts may be
of order the orbital velocity ($v\sim 6,000$ km/s at $\sim 1,000$
Schwarzschild radii), which could be detectable either in individual
objects, or else statistically for the population.

\citet{boroson09} recently reported a candidate SMBH binary, with two
sets of broad emission lines separated by $3,500 {\rm km~s^{-1}}$.
The spectrum of this source can also be interpreted with a single
BH+disk system \citep{halpern88,gaskell09}; indeed, this
interpretation is favored by the lack of any change in the velocity
offset over the course of $\approx 1$ year \citep{chornock09}.
Nevertheless, it is interesting to note that, with the binary
parameters reported for this source (assuming random orientation, and
an expected orbital speed of $\approx 6,000 {\rm km~s^{-1}}$),
$M\approx M_1=10^{8.9}~{\rm M_\odot}$, $M_2=10^{7.3}~{\rm M_\odot}$,
$q=0.025$, $t_{\rm orb}\approx 100$ years, and $R\approx 10^3 R_S$, we
find the evolutionary track of the proposed system to be virtually
indistinguishable from the $M=10^9~{\rm M_\odot}$, $q=0.01$ case shown
in Figure~\ref{fig:tres_r_q0.01}. At its currently observed orbital
separation of $R\approx 10^3 R_S$, the binary would be in the
gas--driven regime, close to outer radius of the formally
gravitationally stable disk (i.e., the system is just outside the
marked $Q=1$ point in this figure), with a residence time of $\approx
10^7$ years. This could indeed make this observed separation common
among quasars with $M=10^9~{\rm M_\odot}$ SMBHBs.  The figure also
shows that the residence time at fixed orbital velocity decreases
steeply with BH mass, suggesting that fainter binary quasars with
similar orbital speeds would be much less common.

In the last stages of coalescence, the GWs emitted by such a system
would induce periodic modulations in the arrival times of pulses from
background radio pulsars; at 200ns timing sensitivity, these
modulations would be detectable from SMBHBs out to a distance of $\sim
20$Mpc \citep[][note that this study already applies the idea to the
source 3C66B mentioned in \S~1, whose elliptical motion was
interpreted as due to a SMBHBs, and rules out the SMBHB
hypothesis]{pulsars}.

\section{Summary and Conclusions}
\label{sec:summary}

In this paper, we followed the evolution of SMBH binaries, starting
from large separations, to coalescence.  We find, in agreement with
earlier works, that the orbital decay is initially generically driven
by viscous binary--disk interactions, whereas GWs dominate the last
stages.  In a refinement of earlier results, we also find that just
prior to the transition to GW--driven evolution, the viscous orbital
decay is generically in the ``secondary--dominated'' Type II migration
regime (the mass of the secondary is larger than the enclosed disk
mass). This is slower than the disk--dominated Type II migration that
has sometimes been assumed in the past, and, as a result, SMBH
binaries spend a significant fraction of their time at orbital periods
of $\sim$days to $\sim$ a year, where they may not be rare, and may be
identifiable.  We emphasized the large uncertainties in the residence
times in this regime -- for example, time--dependent disk models
predict even slower decay.  We also find that observations of BHs with
a mass range of $10^6-10^9~{\rm M_\odot}$ over this range of periods
could find binaries located in all three physically distinct regions
of the circumbinary disk. Thus, several aspect of disk physics could
potentially also be probed in future observations of a population of
SMBH binaries.  We also find that viscous processes may contribute to
the orbital decay rate even after the binaries enter {\it LISA}'s
frequency range, for low-- and/or unequal--mass binaries ($M\lsim
10^5~{\rm M_\odot}$ or $q\lsim 0.01$).  While viscous processes are
strongly sub--dominant for rapidly evolving ``inspiral'' sources,
detected during the last few years of their coalescence, the presence
of the gaseous disk could reduce any background of unresolved
stationary sources at frequencies near the low--frequency end of the
{\it LISA} range ($\sim 10^{-4}$ mHz).

We considered the possibility that there may be a one--to--one
correspondence between the activation of luminous AGN and SMBH
coalescences, with a fraction of AGN exhibiting periodic flux
variations.  Given that the interpretation of individual SMBHB
candidates have so far remained ambiguous, we proposed that a
statistically large sample should aid in the identification of these
binary BH sources.  Our main conclusion is that future surveys in
optical and X--ray bands, which can be sensitive to periodic
variations in the emission from $\sim 10^6-10^9~{\rm M_\odot}$
supermassive black hole binaries, on timescales of $t_{\rm var}\sim$
tens of weeks, at the level of $\sim 1$\% of the Eddington luminosity,
could look for a population of such sources, with the aim of
determining the fraction $f_{\rm var}$ of sources, at a given redshift
and luminosity, as a function of $t_{\rm var}$.  In our simplified
models for the binary--disk interaction, this time--scale of tens of
weeks corresponds to the orbital time when binaries with $M\sim
10^7~{\rm M_\odot}$ make their transition from viscous to GW--driven
evolution. In the latter regime, for sources with $M\gsim 10^7~{\rm
  M_\odot}$, gravitational radiation predicts the scaling $f_{\rm
  var}\propto t_{\rm var}^{8/3}$.  The discovery of a population of
periodic sources whose abundance obeys this scaling would confirm that
the orbital decay is indeed driven by GWs, and also that circumbinary
gas is present at small orbital radii and is being perturbed by the
BHs. Deviations from the $t_{\rm var}^{8/3}$ power--law for
lower--mass BHs would constrain the structure of the circumbinary gas
disk and viscosity--driven orbital decay.

There is certainly a possibility that the periodic sources envisioned
here do not exist (e.g., because the SMBH binary does not produce
bright and variable emission during its GW--emitting stage, at orbital
separations of $\sim 10^3$ Schwarzschild radii).  Nevertheless, we
argued that existing surveys already approach the required combination
of sky coverage and depth, and future surveys, designed to make
observations for several years, with a sampling rate of a few days,
could yield a positive detection and identify periodic source
populations.  This would bring rich scientific rewards, possibly
including the indirect detection of gravitational waves, driving the
orbital decay of these sources.

\acknowledgments

ZH thanks George Djorgovski and Tuck Stebbins for stimulating
discussions, and Mamoru Doi and Tomoki Morokuma for sharing their
Subaru variability search results prior to publication, which
originally inspired this paper.  We also thank Zsolt Frei and David
Hogg for useful comments, Chris Stubbs, Michael Strauss and Richard
Mushotzky for advice on variability surveys, and the anonymous referee
for comments that significantly improved this paper. KM thanks the
Aspen Center for Physics, where a part of the work reported here was
performed, for their hospitality.  This work was supported by the
Pol\'anyi Program of the Hungarian National Office for Research and
Technology (NKTH) and by NASA ATFP grant NNX08AH35G. BK acknowledges
support from OTKA Grant 68228.

\begin{appendix}

\end{appendix}


\clearpage
\newpage
\begin{thebibliography}{10}


\bibitem[Armitage(2007)]{arm07} Armitage, P. J. 2007, preprint arXiv:astro-ph/0701485

\bibitem[Armitage \& Natarajan(2002)]{an02} Armitage, P. J., \& Natarajan, P. 2002, \apj, 567, L9

\bibitem[Armitage \& Natarajan(2005)]{an05} Armitage, P. J., \& Natarajan, P. 2005, \apj, 634, 921

\bibitem[Artymowicz \& Lubow(1994)]{al94} Artymowicz, P., \& Lubow, S. H. 1994, ApJ, 421, 651

\bibitem[Artymowicz \& Lubow(1996)]{al96} Artymowicz, P., \& Lubow, S. H. 1996, ApJ, 467, L77

\bibitem[Bardeen \& Peterson(1975)]{bp75} Bardeen, J. M., \& Peterson, J. A. 1975, ApJ, 195, L65

\bibitem[Barger et al.(2005)]{barger05} Barger, A. J., Cowie, L. L., Mushotzky, R. F., Yang, Y., Wang, W.-H., Steffen, A. T., \& Capak, P. 2005, ApJ, 129, 578

\bibitem[Barnes(2002)]{barnes} Barnes, J. E. 2002, MNRAS, 333, 481

\bibitem[Barnes \& Hernquist(1992)]{bh92} Barnes, J. E., \& Hernquist, L. 1992, \araa, 30, 705

\bibitem[Bauer et al.(2003)]{bauer03} Bauer, F. E., et al. 2003, AN, 324, 175

\bibitem[Bauer et al.(2004)]{bauer04} Bauer, F. E., et al. 2004, AdSpR, 34, 2555

\bibitem[Beckmann et al.(2007)]{beckmann} Beckmann, V., et al. 2007, A\&A, 475, 827

\bibitem[Begelman, Blandford, \& Rees(1980)]{bbr80} Begelman, M. C.,  Blandford, R. D., \&  Rees, M. J. 1980, Nature, 287, 307

\bibitem[Blanchet, Qusailah \& Will(2005)]{blanchet05} Blanchet, L., Qusailah, M. S. S., \& Will, C. M., 2005, ApJ, 635, 508

\bibitem[Bogdanovic, Eracleous \& Sigurdsson(2009)]{bogdanovic09} Bogdanovic, T., Eracleous, M., \& Sigurdsson, S. 2009, \apj, in press, preprint arXiv:0809.3262

\bibitem[Boroson \& Lauer(2009)]{boroson09} Boroson, T. A., \& Lauer, Tod R. 2009, Nature, 458, 53

\bibitem[Chornock et al.(2009)]{chornock09} Chornock, R., Bloom, J. S., Cenko, S. B. et al. 2009, The Astronomer's Telegram, \#1955, http://www.astronomerstelegram.org/?read=1955

\bibitem[Ciotti \& Ostriker(2001)]{co01} Ciotti, L., \& Ostriker, J. P. 2001, ApJ, 551, 131

\bibitem[Corasaniti et al.(2006)]{alpaca} Corasaniti, P. S., LoVerde, M., Crotts, A., \& Blake C. 2006, MNRAS, 369, 798

\bibitem[Crowder \& Cornish(2004)]{cornish04} Crowder, J., \& Cornish, N. J. 2004, \prd, 70, 082004

\bibitem[Cuadra et al.(2009)]{cuadra09} Cuadra, J., Armitage, P. J., Alexander, R. D., \& Begelman, M. C. 2009, 393, 1423

\bibitem[Deffayet \& Menou(2007)]{dm07} Deffayet, C. \& Menou, K. 2007, \apj, 668, L143

\bibitem[Dobbs-Dixon et al.(2007)]{dobbs07} Dobbs-Dixon, I., Li, S. L., \& Lin, D. N. C. 2007, ApJ, 660, 791

\bibitem[Dotti et al.(2006)]{dot06} Dotti, M., Salvaterra, R., Sesana, A., Colpi, M., \& Haardt, F. 2006, \mnras, 372, 869

\bibitem[Dotti et al.(2008)]{dot08} Dotti, M., Colpi, M., Haardt, F., \& Mayer, L. 2008, arXiv eprint, arXiv:0807.3626

\bibitem[Dunkley et al.(2009)]{wmap} Dunkley, J., et al. 2009, \apjs, 180, 306

\bibitem[Escala et al.(2004)]{elcm04} Escala, A., Larson, R. B., Coppi, P. S., \& Mardones, D. 2004, \apj, 607, 765

\bibitem[Escala et al.(2005)]{elcm05} Escala, A., Larson, R. B., Coppi, P. S., \& Mardones, D. 2005, \apj, 630, 152

\bibitem[Fan(2005)]{fanjh05} Fan, J. H. 2005, ChJA\&A, 5, 213

\bibitem[Ferrarese(2002)]{ferrarese02} Ferrarese, L. 2002, ApJ, 578, 90

\bibitem[Ferrarese \& Ford(2005)]{ff05} Ferrarese, L., \& Ford, H. 2005, SSRv, 116, 523

\bibitem[Frank et al.(2002)]{accretionpower} {Frank}, J., {King}, A., \& {Raine}, D.~J. ``Accretion Power in Astrophysics: Third Edition,'' 2002, Cambridge University Press, ISBN 0521620538

\bibitem[Frey et al.(2008)]{frey} Frey, S., Gurvits, L. I., Paragi,  Z., \& Gabanyi, K. E. 2008, A\&A, 484, L39

\bibitem[Fuhrmeister \& Schmitt(2003)]{fuhrmeister} Fuhrmeister, B., \& Schmitt, J. H. M. M. 2003, A\&A, 403, 247

\bibitem[Gaskell et al.(2006)]{variabilitybook} Gaskell, C. M., McHardy, I. M., Peterson, B. M., \& Sergeev, S. G. 2006, ``AGN Variability from X-rays to Radio Waves'', ASP Conf. Series, volume 360

\bibitem[Gaskell(2009)]{gaskell09} Gaskell, C. M. 2009, Nature, submitted, arXiv:0903.4447

\bibitem[Goodman (2003)]{goodman03} Goodman, J. 2003, \mnras 339, 937

\bibitem[Goodman \& Tan (2004)]{goodmantan04} Goodman, J., \& Tan, J. C. 2004, \apj, 608, 108

\bibitem[Gould \& Rix(2000)]{gr00} Gould, A., \& Rix, H.-W. 2000, \apj, 532, 29

\bibitem[Halpern \& Filippenko(1988)]{halpern88} Halpern, J. \& Filippenko, A. 1988, Nature, 331, 46

\bibitem[Hayasaki(2009)]{hayasaki09} Hayasaki, K. 2009, \pasj, in press, e-print arXiv:0805.3408

\bibitem[Hayasaki et al.(2008)]{hayasaki08} Hayasaki, K., Mineshige, S., \& Ho, L. C. 2008, \apj, 682, 1134

\bibitem[Hayasaki et al.(2007)]{hayasaki07} Hayasaki, K., Mineshige, S., Sudou, H. 2007, \pasj, 59, 427

\bibitem[Hirose, Krolik \& Blaes(2009)]{hirose09} Hirose, S., Krolik, J. H., \& Blaes, O. 2009, \apj, 691, 16

\bibitem[Holz \& Hughes(2005)]{hh05} Holz, D. E., \& Hughes, S. A. 2005, \apj, 629, 15

\bibitem[Hopkins et al.(2005)]{hopkins05} Hopkins, P. F., Hernquist, L., Cox, T. J., Di Matteo, T., Robertson, B., \&  Springel, V. 2005, ApJ, 630, 716

\bibitem[Hopkins et al.(2006)]{hopkins06} Hopkins, P. F., Hernquist, L., Cox, T. J., Di Matteo, T., Robertson, B., \&  Springel, V. 2006, ApJS, 163, 1

\bibitem[Hopkins et al.(2007a)]{hopkins07a} Hopkins, P. F., Bundy, K., Hernquist, L., \& Ellis, R. S. 2007a, ApJ, 659, 976

\bibitem[Hopkins et al.(2007b)]{hopkins07b} Hopkins, P. F., Richards, G. T., \& Hernquist, L. 2007b, ApJ, 654, 731


\bibitem[Hughes(2007)]{hughes07} Hughes, S. A. 2007, in the Proceedings of the 7th Edoardo Amaldi Conference on Gravitational Waves (to be published by Classical and Quantum Gravity), in press, e-print arXiv:0711.0188

\bibitem[Ivanov, Papaloizou, \& Polnarev(1999)]{ivanov} Ivanov, P. B., Papaloizou, J. C. B., \& Polnarev, A. G. 1999, \mnras, 307, 79

\bibitem[Jenet et al.(2004)]{pulsars} Jenet, F. A., Lommen, A., Larson, S. L., \& Wen L. 2004, ApJ, 606, 799

\bibitem[Kauffman \& Haehnelt(2000)]{kh00}  Kauffmann, G., \& Haehnelt, M. 2000, \mnras, 311, 576

\bibitem[Kazantzidis et al.(2005)]{kazantzidis05} Kazantzidis, S., et~al. 2005, \apjl, 623, L67

\bibitem[Kocsis et al.(2006)]{koc06} Kocsis, B., Frei, Z., Haiman, Z. \& Menou, K. 2006, \apj , 637, 27

\bibitem[Kocsis et al.(2007)]{paper1} Kocsis, B., Haiman, Z., Menou, K., \& Frei, Z. 2007, \prd, 76, 022003

\bibitem[Kocsis et al.(2008)]{paper2} Kocsis, B., Haiman, Z., \& Menou, K. 2008, ApJ, 684, 870

\bibitem[Kocsis \& Loeb(2008)]{kocsisloeb08}  Kocsis, B., \& Loeb, A., 2008, \prl, 101, 041101

\bibitem[Kollmeier et al.(2006)]{kollmeier06} Kollmeier, J. A., et al. 2006, ApJ, 648, 128

\bibitem[Komossa et al.(2003)]{komossa03} Komossa, S. et al. 2003, \apj, 582, L15

\bibitem[Komossa, Zhou \& Lu(2008)]{komossa08} Komossa, S., Zhou, H., \& Lu, H. 2008, \apj, 678, L81

\bibitem[Kormendy \& Richstone(1995)]{kr95} Kormendy, J., \& Richstone, D.\ 1995, \araa, 33, 581

\bibitem[Lang \& Hughes(2006)]{lh06} Lang, R. \& Hughes, S. A. 2006, \prd, 74, 122001

\bibitem[Lang \& Hughes(2008)]{lh08} Lang, R. N., \& Hughes, S. A. 2008, \apj, 677, 1184

\bibitem[Lightman \& Eardley(1974)]{le74} Lightman, A. P., \& Eardley, D. M. 1974, \apj, 187, L1

\bibitem[Lippai, Frei \& Haiman(2008a)]{lippaidisk} Lippai, Z., Frei, Zs., \& Haiman, Z. 2008a, ApJ, 676, L5

\bibitem[Lippai, Frei \& Haiman(2008b)]{lippaimergers} Lippai, Z., Frei, Zs., \& Haiman, Z. 2008b, ApJ, submitted

\bibitem[Liu(2004)]{liu04} Liu, F. K. 2004, MNRAS,347, 1357

\bibitem[Liu, Liu \& Xie(1997)]{liu97} Liu, F. K., Liu, B. F., \& Xie, G. Z. 1997, A\&AS, 123, 569

\bibitem[Liu, Wu \& Cao(2003)]{liu03} Liu F. K., Wu  X.-B., \& Cao S.L. 2003, MNRAS, 340, 411

\bibitem[Liu, Xie \& Bai(1995)]{liu95} Liu, F. K., Xie, G. Z., \& Bai, J. M. 1995, A\&A, 295, 1

\bibitem[Loeb(2007)]{loeb07} Loeb, A. 2007, PRL, 99, d1103

\bibitem[Lynden-Bell \& Pringle(1974)]{lp74} Lynden-Bell, D., \& Pringle, J. E. 1974, \mnras, 168, 603

\bibitem[MacFadyen \& Milosavljevi\'c(2008)]{mm08} MacFadyen, A., \& Milosavljevi\'c, M. 2008, ApJ, 672, 83

\bibitem[MacLeod \& Hogan(2007)]{mh07} MacLeod, C. L., \& Hogan, C. J. 2007, Phys. Rev. D., vol. 77, Issue 4, id. 043512

\bibitem[Markowitz \& Edelson(2001)]{markowitz} Markowitz, A., \& Edelson, R. 2001, ApJ, 547, 684

\bibitem[Martini(2004)]{martini04} Martini, P.  2004, in Carnegie Observatories Astrophysics Series, Vol. 1: Coevolution of Black Holes and Galaxies, ed. L. C. Ho (Cambridge: Cambridge Univ. Press), p. 169

\bibitem[Mayer, Kazantzidis, \& Escala(2008)]{mke08} Mayer, L., Kazantzidis, S., \& Escala, A. 2008, preprint arXiv:0807.3329

\bibitem[Menou, Haiman \& Narayanan(2001)]{mhn01}Menou, K., Haiman,  Z., \& Narayanan, V. K. 2001, \apj, 558, 535

\bibitem[Menou \& Quataert(2001)]{mq01} Menou, K., \& Quataert, E. 2001, ApJ, 552, 204

\bibitem[Merritt \& Ekers(2002)]{merritt} Merritt, D.,  Ekers, R. D. 2002, Science 297, 1310

\bibitem[Milosavljevic \& Phinney(2005)]{mp05} Milosavljevic, M., \& Phinney, E. S. 2005, \apj, 622, L93

\bibitem[Morokuma et al.(2008a)]{morokuma1} Morokuma et al. 2008a, ApJ, 676, 163

\bibitem[Morokuma et al.(2008b)]{morokuma2} Morokuma et al. 2008b, ApJ, 676, 121

\bibitem[Neugebauer \& Matthews(1999)]{neugebauer} Neugebauer, G., \& Matthews, K. 1999, AJ, 118, 35

\bibitem[Padmanabhan(2002)]{padmanabhan} Padmanabhan, T., ``Theoretical Astrophysics,'' 2002, Cambridge University Press, ISBN 0521562422, Vol. 1, Eq. 6.235

\bibitem[Qian et al.(2007)]{qian07} Quian, S.-J., et al. 2007, ChJA\&A, 7, 364

\bibitem[Papadakis et al.(2008)]{papadakis08} Papadakis, I. E., Chatzopoulos, E., Athanasiadis, D., Markowitz, A., \& Georgantopoulos, I. 2008, A\&A, 487, 475

\bibitem[Paolillo et al.(2004)]{paolillo04} Paolillo, M., Schreier, E. J., Giacconi, R., Koekemoer, A. M. \& Grogin, N. A. 2004, ApJ, 611, 93

\bibitem[Rafikov(2002)]{rafikov02} Rafikov, R. R. 2002, ApJ, 572, 566

\bibitem[Raiteri et al.(2001)]{raiteri01} Raiteri, C. M., et al. 2001, A\&A, 377, 396

\bibitem[Rieger(2007)]{rieger07} Rieger, F. M. 2007, Ap\&SS, 309, 271

\bibitem[Robertson et al.(2006)]{robertson06} Robertson, B., Cox, T. J., Hernquist, L, Franx, M., Hopkins, P. F., Martini, P., \& Springel, V. 2006, ApJ, 641, 21

\bibitem[Rodriguez et al.(2006)]{rodriguez06} Rodriguez, C. et al. 2006, ApJ, 646, 49

\bibitem[Roos, Kaastra \& Hummel(1993)]{roos} Roos, N.,  Kaastra, J. S., \& Hummel, C. A. 1993, ApJ, 409, 130

\bibitem[Rybicki \& Lightman(1986)]{rybicki} {Rybicki}, G. B., \& {Lightman}, A. P., ``Radiative Processes in Astrophysics,'' 1986, Wiley-VCH, ISBN 0-471-82759-2, Eq. 5.20

\bibitem[Sarajedini(2008)]{sarajedini} Sarajedini, V. L. 2008, Rev. Mex. A\&A, 32, 34

\bibitem[Sarajedini et al.(2006)]{sarajedini06} Sarajedini, V. L., et al. 2006, ApJS, 155, 271

\bibitem[Schoenmakers(2000)]{schoenmakers} Schoenmakers, A. P. 2000, MNRAS, 315, 371

\bibitem[Schutz(1986)]{sch86} Schutz, B. F. 1986, Nature, 323, 310

\bibitem[Schnittman et al.(2008)]{anatomy08} Schnittman, J. D., Buonanno, A., van Meter, J. R., Baker, J. G., Boggs, W. D., Centrella, J., Kelly, B. J., \& McWilliams, S. T. 2008, Phys. Rev. D., vol. 77, Issue 4, id. 044031

\bibitem[Schnittman \& Krolik(2008)]{schnittmankick} Schnittman, J. D., \& Krolik, J. H., 2008, \apj, 684, 835

\bibitem[Sesana et al.(2004)]{sesana04} Sesana, A., Haardt, F., Madau, P., \& Volonteri, M. 2004, \apj, 611, 623

\bibitem[Sesana et al.(2005)]{sesana05} Sesana, A., Haardt, F., Madau, P., \& Volonteri, M. 2005, \apj, 623, 23

\bibitem[Shakura \& Sunyaev(1973)]{ss73} Shakura, N. I., \& Syunyaev, R. A. 1973, A\&A, 24, 337

\bibitem[Shapiro \& Teukolsky(1986)]{shapiroteukolsky} Shapiro, S. L., \& Teukolsky, S. A., ``Black Holes, White Dwarfs and Neutron Stars: The Physics of Compact Objects,'' 1986, Wiley-VCH, ISBN 0-471-87316-0, Eq. 14.5.26

\bibitem[Shields \& Bonning(2008)]{shieldskick} Shields, G. A., \& Bonning, E. W., 2008, \apj, 682, 758

\bibitem[Sillanp\"{a}\"{a} et al.(1988)]{sillanpaa} Sillanp\"{a}\"{a}, A., Haarala, S., Valtonen, M. J., Sundelius, B. \& Byrd, G. G. 1988, \apj, 325, 628

\bibitem[Sirko \& Goodman(2003)]{sirkogoodman03} Sirko, E., \& Goodman, J. 2003, \mnras, 341, 501

\bibitem[Springel, Di Matteo \& Hernquist(2005)]{springel05} Springel, V., Di Matteo, T., \& Hernquist, L. 2005, ApJ, 620, L79

\bibitem[Stubbs(2008)]{stubbs} Stubbs, C. W. 2008, in the Proceedings of the 12th Gravitational Wave Data Analysis Workshop, to appear in Classical and Quantum Gravity, preprint arXiv:0712.2598

\bibitem[Sudou et al.(2003)]{sudou} Sudou, H.,  Iguchi, S., Murata, Y.,  Taniguchi,  Y. 2003, Science, 300, 1263

\bibitem[Syer \& Clarke(1995)]{syerclarke}	Syer, D., \& Clarke, C. J. 1995, \mnras, 277, 758

\bibitem[Tanaka, Takeuchi, \& Ward(2002)]{tanaka02} Tanaka, H., Takeuchi, T., \& Ward, W. R. 2002, \apj, 565, 1257

\bibitem[Tao et al.(2008)]{tao08} Tao, J., Fan, J., Qian, B., \& Liu, Y. 2008, AJ, 135, 737

\bibitem[Vanden Berk et al.(2004)]{vandenberk04} Vanden Berk, D. E. et al. 2004, \apj, 601, 692

\bibitem[Vecchio(2004)]{vec04} Vecchio, A. 2004, \prd, 70, 042001

\end{thebibliography}
\end{document}